\documentclass[aps,prl,preprint,showpacs,showkeys,superscriptaddress]{revtex4}
\usepackage{pstricks}
\usepackage{epsfig}
\usepackage{graphicx}
\usepackage{amsmath}
\usepackage{amsfonts}
\usepackage{amssymb}
\usepackage{wasysym}
\usepackage{bbold}
\usepackage{rotating}
\newcommand{\bra}[1]{\langle #1|}
\newcommand{\ket}[1]{|#1\rangle}
\newcommand{\braket}[2]{\langle #1|#2\rangle}

\usepackage{multirow}%multiple rows in a table are represented by one row
\begin{document}
\title{Extensive v2DM study of the one-dimensional Hubbard model for large lattice sizes: Exploiting translational invariance and parity}
\author{Brecht Verstichel}
\email{brecht.verstichel@ugent.be}
\affiliation{Ghent University, Center for Molecular Modeling, Technologiepark 903, 9052 Zwijnaarde, Belgium}
\author{Helen van Aggelen}
\affiliation{Ghent University, Department of Inorganic and Physical Chemistry, Krijgslaan 281 (S3), B-9000 Gent, Belgium}
\author{Ward Poelmans}
\affiliation{Ghent University, Center for Molecular Modeling, Technologiepark 903, 9052 Zwijnaarde, Belgium}
\author{Sebastian Wouters}
\affiliation{Ghent University, Center for Molecular Modeling, Technologiepark 903, 9052 Zwijnaarde, Belgium}
\author{Dimitri Van Neck}
\affiliation{Ghent University, Center for Molecular Modeling, Technologiepark 903, 9052 Zwijnaarde, Belgium}
\begin{abstract}
Using variational density matrix optimization with two- and three-index conditions we study the one-dimensional Hubbard model with periodic boundary conditions at various filling factors. Special attention is directed to the full exploitation of the available symmetries, more specifically the combination of translational invariance and space-inversion parity, which allows for the study of large lattice sizes. We compare the computational scaling of three different semidefinite programming algorithms with increasing lattice size, and find the boundary point method to be the most suited for this type of problem. Several physical properties, such as the two-particle correlation functions, are extracted to check the physical content of the variationally determined density matrix. It is found that the three-index conditions are needed to correctly describe the full phase diagram of the Hubbard model. We also show that even in the case of half filling, where the ground-state energy is close to the exact value, other properties such as the spin-correlation function can be flawed.
\end{abstract}
\maketitle
\section{Introduction}
%Hier wat over Coleman enzo.

The reduced density matrix makes its first appearance in the work of Dirac, in which the single-particle density matrix (1DM) is used in the description of Hartree-Fock theory \cite{dirac_dm}. Husimi \cite{husimi} was the first to note that, for a system of identical particles interacting only in a pairwise manner, the energy can be expressed exactly as a function of the 2DM. This becomes very clear in second-quantized notation (see {\it e.g.} \cite{fetter_walecka,bijbel}), where a system of identical particles interacting pairwise is described by a general Hamiltonian:
\begin{equation}
\hat{H} = \sum_{\alpha\beta}t_{\alpha\beta}a^\dagger_\alpha a_\beta + \frac{1}{4}\sum_{\alpha\beta\gamma\delta}V_{\alpha\beta;\gamma\delta}a^\dagger_\alpha a^\dagger_\beta a_\delta a_\gamma~.
\label{2pham}
\end{equation}
The expectation value for the energy of any ensemble of $N$-particle wave functions $\ket{\Psi_i^N}$ with positive weights $w_i$, can then be expressed as a function of the 2DM alone:
\begin{equation}
\sum_iw_i\bra{\Psi^N_i}\hat{H}\ket{\Psi^N_i}= \mathrm{Tr}~\Gamma H^{(2)} = \frac{1}{4}\sum_{\alpha\beta\gamma\delta}\Gamma_{\alpha\beta;\gamma\delta}H^{(2)}_{\alpha\beta;\gamma\delta}~,
\label{ener_func}
\end{equation}
in which we have introduced the 2DM:
\begin{equation}
\Gamma_{\alpha\beta;\gamma\delta} = \sum_iw_i\bra{\Psi^N_i}a^\dagger_\alpha a^\dagger_\beta a_\delta a_\gamma \ket{\Psi^N_i}~,\qquad\text{with}\qquad\sum_iw_i=1~,
\label{2DM_intro}
\end{equation}
and the reduced two-particle Hamiltonian,
\begin{equation}
H^{(2)}_{\alpha\beta;\gamma\delta} = \frac{1}{N-1}\left(\delta_{\alpha\gamma}t_{\beta\delta} - \delta_{\alpha\delta}t_{\beta\gamma} - \delta_{\beta\gamma}t_{\alpha\delta} + \delta_{\beta\delta}t_{\alpha\gamma}\right) + V_{\alpha\beta;\gamma\delta}~.
\label{reduced_ham}
\end{equation}
The idea to use the 2DM as a variable in a variational scheme was first published in literature by L\"owdin in his groundbreaking article \cite{lowdin}, but even earlier, in 1951, John Coleman tried a practical variational calculation on Lithium. To his surprise, the energy he obtained was far too low, after which he realized the variation was performed over too large a class of 2DM's \cite{rdm_book}. Independently and unaware of the work by L\"owdin and Coleman, Joseph Mayer \cite{mayer} used the 2DM in a study of the electron gas. In a reply to Mayer's paper, Tredgold \cite{tredgold} pointed out the unphysical nature of the results, and suggested that additional conditions on the density matrix are needed to improve on them. 

These results led Coleman, in his seminal review paper \cite{coleman}, to formulate the $N$-rep\-re\-sen\-ta\-bi\-li\-ty problem. This is the problem of finding the necessary and sufficient conditions which a reduced density matrix has to fulfil to be derivable from a statistical ensemble of physical wave functions, {\it i.e.} expressible as in Eq.~(\ref{2DM_intro}). In this paper he also derived the necessary and sufficient conditions for ensemble $N$-representability of the 1DM, and some bounds on the eigenvalues of the 2DM. A big step forward was the derivation of the $\mathcal{Q}$ and $\mathcal{G}$ matrix non-negativity conditions by Garrod and Percus \cite{garrod}. These were practical constraints, which allowed for a computational treatment of the problem. The first numerical calculation using these conditions on the Beryllium atom \cite{fusco,garrod_mih_ros} was very encouraging, as the results were highly accurate. It turned out, however, that Beryllium, due to its simple electronic stucture, is a special case where these conditions perform very well. A subsequent study showed that these conditions do not work well at all for nuclei \cite{mihailovic,rosina}. This disappointing result, together with the computational complexity of the problem, caused activity in the field to diminish for the next 25 years. The change came with the development of a new numerical technique, called semidefinite programming, which turned out to be very suited for the determination of the 2DM under matrix non-negativity constraints. Maho Nakata {\it et al.} \cite{nakata_first} were the first to use a standard semidefinite programming package to calculate the ground-state energies of some atoms and molecules, and obtained quite accurate results. He was quickly followed by the extensive work of Mazziotti \cite{mazziotti}. These results reinvigorated interest in the method, and sparked of a lot of developments. New $N$-representability conditions were introduced, {\it e.g.} the three-index $\mathcal{T}$ conditions, set forth by Zhao {\it et al.} \cite{zhao}, which led to mHartree accuracy \cite{hammond,nakata_last,mazz_T_con,Gido_T_con,mazz_book,braams_book} for molecules near equilibrium geometries. 

In recent years interest in the method has been growing, as the variational determination of the 2DM results in a lower bound, which is highly complementary to the upper bound obtained in variational approaches based on a wave-function ansatz. In addition, the method is essentially non-perturbative in nature, and has a completely different structure quite unrelated to other many-body techniques. A lot of activity has been devoted to the search for new $N$-representability conditions, which improve the result in a computationally cheap way \cite{dimi,qsep}. There have also been efforts to improve the semidefinite programming algorithms by adapting them to the specific problem of density matrix optimization \cite{maz_prl,maz_bp,primal_dual}, allowing the study larger systems.

The one-dimensional Hubbard model \cite{hubbard} is the simplest model possessing non-trivial correlations present in a solid. The Hamiltonian reads:
\begin{equation}
\hat{H} = -t\sum_{i\sigma}\left(a^\dagger_{i;\sigma}a_{i+1;\sigma} + a^\dagger_{i+1;\sigma}a_{i;\sigma}\right) + U\sum_{i}a^\dagger_{i\uparrow}a_{i\uparrow}a^\dagger_{i\downarrow}a_{i\downarrow}~.
\label{hubbard}
\end{equation}
It pertains to a one-dimensional lattice, with sites labeled by $i = 1,\ldots,L$. Periodic boundary conditions (PBC) are assumed.

\begin{figure}
\centering
\includegraphics[scale=0.5]{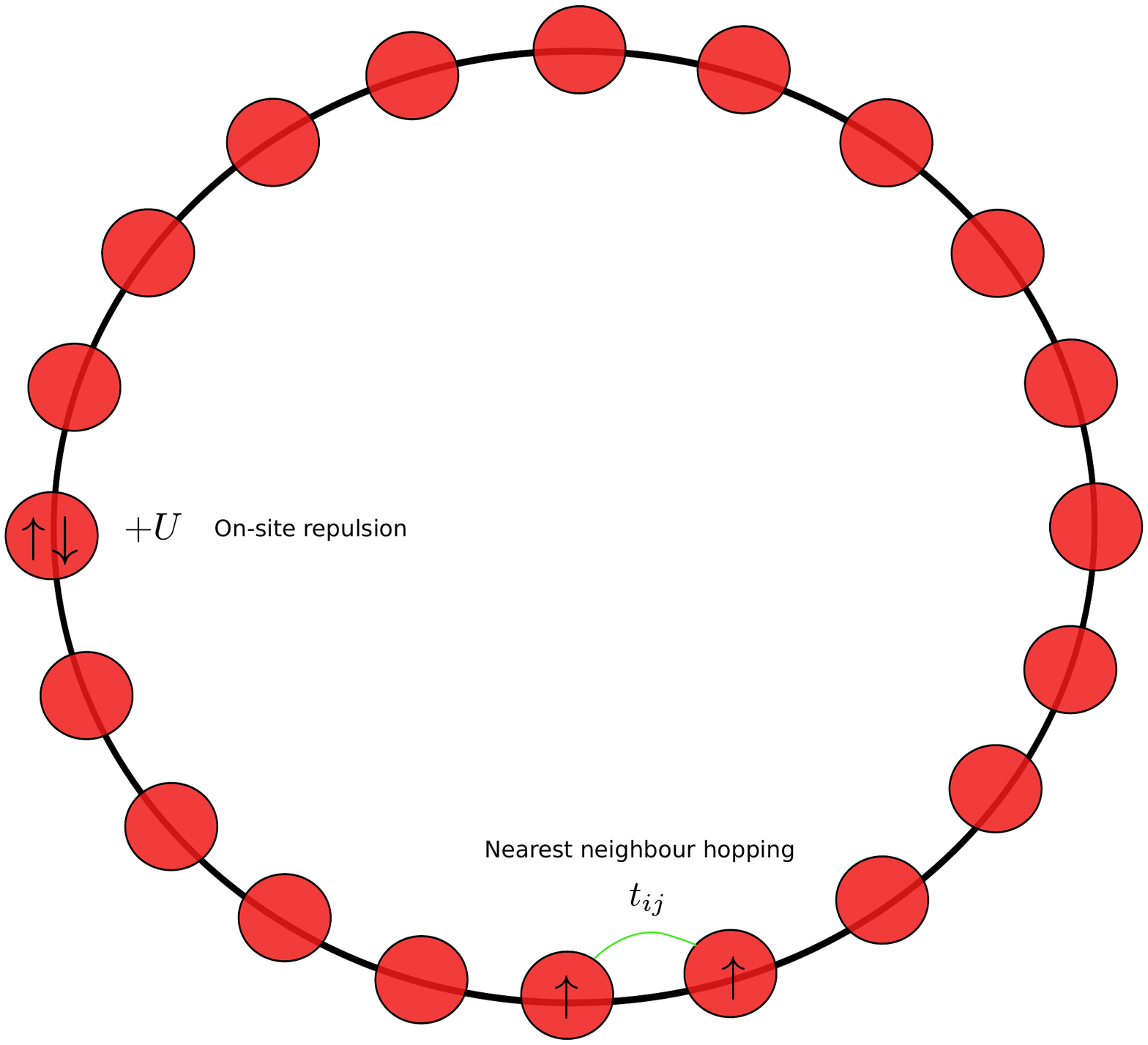}
\caption{\label{hubbard_fig} Illustration of the two terms present in the Hubbard Hamiltonian, electrons can jump to nearest-neighbour sites with amplitude $t_{ij}$. When two electrons are on the same site, there is an energy penalty of $U$.}
\end{figure}

The complexity of this seemingly simple schematic Hamiltonian lies in the competition between the first term, called the hopping term, which delocalizes the electrons, and the second on-site repulsion term which is diagonal in the site basis. In Figure \ref{hubbard_fig} a graphic representation of the two terms in the Hubbard Hamiltonian is shown. 

In this article the one-dimensional Hubbard is studied using the standard $\mathcal{IQG}$ two-index and $\mathcal{T}_{1,2}$ three-index non-negativity conditions. In the next Section we discuss how the huge amount of symmetry which is present in the model can be exploited to get a significant speedup of the programs. More specifically, it is shown how translational invariance and space-inversion parity can be combined. In the subsequent Section this symmetry is used to compare the computational scaling of different semidefinite programming algorithms with increasing lattice sites. In the final Section we show the v2DM results using two- and three-index constraints for the one-dimensional Hubbard model on a 20- and 50-site lattice with different filling factors. We have not only computed the ground-state energy, but also compared the spin and charge two-particle correlations functions with Quantum Monte Carlo \cite{sorella} and Bethe ansatz  \cite{ogata} results, to check the validity of the variationally obtained 2DM.

%algemeen, zoals gewoonlijk, en dan symmetrie, translationele invariantie + pariteit
\section{\label{symmetry}Exploiting symmetry in the Hubbard model}
As there is no preferred direction in spin space in Eq.~(\ref{hubbard}), spin symmetry can be exploited and all the results from \cite{atomic} are taken over. However, far more symmetries are present in the model, which allow for a huge reduction of the matrix dimensions involved.
\subsection{Translational invariance}
When periodic boundary conditions are assumed (as in Figure \ref{hubbard_fig}) the Hamiltonian is invariant under translations along the lattice. This is an abelian symmetry for which it is easy to block diagonalize the 2DM and its matrix maps, as the correct basis transformation in single-particle space automatically block diagonalizes all matrices on higher-order particle space. Translational invariance is exploited by Fourier transforming the site basis to quasi-momentum space:
\begin{equation}
\ket{k\sigma} = \sqrt{\frac{1}{L}}\sum_j e^{ikj}\ket{j\sigma}~,
\label{fourier}
\end{equation}
where $L$ is the lattice size, and $k$ takes on the values $\frac{2\pi n}{L}$ for $n = 0,\ldots,L-1$. The kinetic or hopping part of the Hamiltonian becomes diagonal in this basis:
\begin{equation}
\hat{H}_{\text{hop}} = -2t \sum_{k\sigma} \cos{k} ~ a^\dagger_{k\sigma}a_{k\sigma}~,
\label{sp_hubbard}
\end{equation}
from which it follows that in the non-interacting ($U=0$) ground state, the electrons occupy the states with momenta lower than the Fermi level.

The eigenstates of the Hubbard Hamiltonian have a good total quasi-momentum $\mathcal{K}$. The 2DM for these states, expressed in the quasi-momentum single-particle basis:
\begin{equation}
\Gamma^{S}_{k_ak_b;k_ck_d} = \sum_i w_i \frac{1}{[\mathcal{S}]^2} \sum_\mathcal{M}\bra{\Psi^{N\mathcal{K}}_{\mathcal{SM},i}}{B^\dagger}^{S}_{k_ak_b}~B^{S}_{k_ck_d}\ket{\Psi^{N\mathcal{K}}_{\mathcal{SM},i}}~,
\label{2DM_mom}
\end{equation}
where
\begin{equation}
{B^\dagger}^{S}_{k_ak_b} = \frac{1}{\sqrt{1 + \delta_{k_ak_b}}}\left[a^\dagger_{k_a}\otimes a^\dagger_{k_b}\right]^{S}
=\frac{1}{\sqrt{1 + \delta_{k_ak_b}}}\sum_{\sigma_a\sigma_b}\braket{\frac{1}{2}\sigma_a\frac{1}{2}\sigma_b}{SM}a^\dagger_{k_a\sigma_a}a^\dagger_{k_b\sigma_b} ~,
\label{2DM_op_transin}
\end{equation}
is automatically block diagonal, because the only non-zero matrix elements in Eq.~(\ref{2DM_mom}) are those which conserve momentum: $(k_a + k_b)\%2\pi =(k_c + k_d)\%2\pi$\footnote{Here \% signifies the modulo operator}. This means we have $L$ blocks $\Gamma^{SK}$ for every $S$, with two-particle states that satisfy $K = (k_a+k_b)\%2\pi$, and a block dimension that scales linearly with lattice size $L$.

The spin-symmetric matrix constraints simplify considerably by including translational invariance, because the 1DM is automatically diagonal in the quasi-momentum basis:
\begin{equation}
\rho_{k} = \sum_i w_i \frac{1}{[\mathcal{S}]^2} \sum_\mathcal{M}\bra{\Psi^{N\mathcal{K}}_{\mathcal{SM},i}}a^\dagger_{k\sigma}a_{k\sigma}\ket{\Psi^{N\mathcal{K}}_{\mathcal{SM},i}}~.
\label{1DM_ti}
\end{equation}
The translationally invariant 1DM can be derived from the 2DM as:
\begin{equation}
\rho_k = \frac{1}{N-1}\sum_S\frac{[S]^2}{2}\sum_{k'}(1 + \delta_{kk'})\Gamma^{SK}_{kk';kk'}~.
\label{1DM_sc_ti_written_up}
\end{equation}
As an example the translationally invariant form of the $\mathcal{G}$ condition is shown. There is a slight complication because the correct annihilation or hole operator is given by:
\begin{equation}
\tilde{a}_{k\sigma} = (-1)^{\frac{1}{2}+\sigma} a_{\bar{k}-\sigma}~,\qquad\text{with}\qquad \bar{k} = -k\%2\pi~.
\label{good_sto_mom}
\end{equation}
Using this operator the translationally invariant $\mathcal{G}$ map becomes:
\begin{equation}
\mathcal{G}(\Gamma)^{SK}_{k_ak_b;k_ck_d} = \sum_i w_i \frac{1}{[\mathcal{S}]^2}\sum_\mathcal{M} \bra{\Psi^{\mathcal{K}N}_{\mathcal{SM},i}}{A^\dagger}^{SK}_{k_ak_b}~{A}^{SK}_{k_ck_d}\ket{\Psi^{N\mathcal{K}}_{\mathcal{SM},i}}~,
\end{equation}
where $K = (k_a+k_b)\%2\pi = (k_c+k_d)\%2\pi$ and with the particle-hole operator defined by:
\begin{equation}
{A^\dagger}^{SK}_{k_ak_b} = \left[a^\dagger_{k_a}\otimes \tilde{a}_{k_b}\right]^{SK}~.
\end{equation}
The $\mathcal{G}$ map can be expressed as a function of the 2DM:
\begin{equation}
\mathcal{G}(\Gamma)^{SK}_{k_ak_b;k_ck_d} = \delta_{k_bk_d}\delta_{k_ak_c}\rho_{k_a} - \sqrt{( 1 + \delta_{k_a\bar{k}_d})(1 + \delta_{k_b\bar{k}_c})}\sum_{S'}
\left\{
\begin{matrix}
\frac{1}{2}&\frac{1}{2}&S\\
\frac{1}{2}&\frac{1}{2}&S'
\end{matrix}
\right\}
\Gamma^{S'K'}_{k_a\bar{k}_d;k_c\bar{k}_b}~,
\end{equation}
from which one can see that the blocks in the $\mathcal{G}$ matrix with $K = (k_a + k_b)\%2\pi= (k_c + k_d)\%2\pi$, correspond to the blocks with $K' = (k_a +\bar{k}_d)\%2\pi= (k_c +\bar{k}_b)\%2\pi$ in the 2DM, as expected for a particle-hole transformed quantity.

\subsection{Parity}
The Hubbard model with periodic boundary conditions (PBC) has more symmetries than spin and translational invariance, one of them being parity. Parity follows from the symmetry under the inversion of space, {\it i.e.} $\mathbf{r}\rightarrow -\mathbf{r}$. One can readily appreciate from Figure~\ref{hubbard_fig} that this symmetry is present here, {\it i.e.} the Hamiltonian is invariant for the inversion of the site index $i\rightarrow-i\%L$. From the Fourier transform in Eq.~(\ref{fourier}) it can be seen that the effect of this operator on a momentum state is to transform $k$ into $\bar{k} = -k\%2\pi$. However, this operation does not commute with translation, which means that the eigenstates of the Hamiltonian cannot have good momentum \emph{and} parity at the same time. From now on we assume that the number of lattice sites $L$ is even.

As a consequence, if the ground state has momentum $\ket{\mathcal{K}}\neq 0$ or $\pi$ then it is doubly degenerate, forming a doublet with $\ket{\overline{\mathcal{K}}}$. In what follows we will use this degeneracy to exploit both translational invariance and parity to reduce the dimensions of the matrices involved in the program. The following considerations are valid for every $\mathcal{K}$. We start with the simplest case, the 1DM.

Single-particle space is built out of $L$ different momentum states having up or down spin, $\ket{k\sigma}$, for $0 \leq k < 2\pi$ and $\sigma=\pm\frac{1}{2}$. If we transform to a basis with good parity, momentum is no longer a good quantum number:
\begin{equation}
\ket{\tilde{k}^\pi\sigma} = {}^\rho N_k \left(\ket{k\sigma} + \pi\ket{\bar{k}\sigma}\right)~,\qquad\text{with}\qquad 0\leq\tilde{k}\leq\pi~.
\end{equation}
Two states, $k=0$ and $k=\pi$ \footnote{Note that we use $\pi$ for both the parity quantum number, $\pi=\pm1$ as for the transcendent number; in what follows it is always clear from the context what interpretation should be given to $\pi$.}
are mapped on themselves, and only positive parity states can be formed with these momenta. They have norms ${}^\rho N_k = \frac{1}{2}$. For the other states, $0<k<\pi$, both positive and negative parity combinations can be constructed, with norm ${}^\rho N_k=\frac{1}{\sqrt{2}}$. To take advantage of both symmetries at the same time, we define the 1DM using an ensemble of the $\ket{\mathcal{K}}$ and $\ket{\mathcal{\bar{K}}}$ states:
\begin{align}
\rho_{\tilde{k}^\pi\tilde{k}'^{\pi'}} =& \sum_iw_i\frac{1}{2}\frac{1}{[\mathcal{S}]^2}\sum_{\mathcal{M}}\sum_{p\in\pm}\bra{\Psi^{N(p\mathcal{K})}_{\mathcal{SM},i}}a^\dagger_{\tilde{k}^\pi\sigma}a_{\tilde{k}'^{\pi'}\sigma}\ket{\Psi^{N(p\mathcal{K})}_{\mathcal{SM},i}}\\
=&\frac{{}^{\rho}N_k^2}{2}\sum_{p\in\pm}\left[\delta_{kk'}\left(\rho^{p\mathcal{K}}_{k}+\pi\pi'\rho^{p\mathcal{K}}_{\bar{k}}\right) + \delta_{k\bar{k}'}\left(\pi'\rho^{p\mathcal{K}}_{k}+\pi\rho^{p\mathcal{K}}_{\bar{k}}\right)\right]~,
\label{1DM_ti_par}
\end{align}
in which $\rho^{p\mathcal{K}}_k$ is a regular translationally invariant 1DM as defined in Eq.~(\ref{1DM_ti}). Because of parity symmetry we have,
\begin{equation}
\rho^{\mathcal{K}}_k = \rho^{\bar{\mathcal{K}}}_{\bar{k}}~,
\end{equation}
from which it follows that Eq.~(\ref{1DM_ti_par}) is diagonal in $\pi$. If we now define:
\begin{equation}
\rho_k = \frac{1}{2}\left(\rho_k^{\mathcal{K}}+\rho_k^{\bar{\mathcal{K}}}\right)~,
\end{equation}
the translationally invariant 1DM with good parity reduces to:
\begin{equation}
\rho_{\tilde{k}^\pi} = \frac{1}{2}\left(\rho_k + \rho_{\bar{k}}\right)~,
\end{equation}
which can be seen to be independent of parity. The 1DM is still diagonal in $\tilde{k}$, but less elements have to be stored, since for $0<\tilde{k}<\pi$ there is a degeneracy in parity.

In contrast to parity in atomic systems (see {\it e.g.} \cite{atomic}), the parity of a two-particle state for translationally invariant systems is not the product of the parities of the single-particle states building up the two-particle state. Instead, the parity is inherently a two-particle property, and the single-particle states building up the two-particle states have no good parity:
\begin{equation}
\ket{ab;S\tilde{K}^\pi} = {}^{\Gamma}N_{ab}^K\left(\ket{ab;SK} + \pi\ket{\bar{a}\bar{b};S\bar{K}}\right)~,
\label{tp_ti_par}
\end{equation}
with
\begin{equation}
0\leq \tilde{K}\leq\pi\qquad\text{and}\qquad0\leq a,b < 2\pi~.
\end{equation}
In general the 2DM is defined using a $p\mathcal{K}$ ensemble:
\begin{align}
\Gamma^{S\tilde{K}^\pi}_{ab;cd} =& \sum_iw_i\frac{1}{2}\frac{1}{[\mathcal{S}]^2}\sum_{\mathcal{M}}\sum_{p\in\pm}\bra{\Psi^{N(p\mathcal{K})}_{\mathcal{SM},i}}{B^\dagger}^{S\tilde{K}^\pi}_{ab}B^{S\tilde{K}^\pi}_{cd}\ket{\Psi^{N(p\mathcal{K})}_{\mathcal{SM},i}}~,
\end{align}
where
\begin{equation}
{B^\dagger}^{S\tilde{K}^\pi}_{ab} = {}^{\Gamma}N^{K}_{ab}\left({B^\dagger}^{SK}_{ab}+\pi{B^\dagger}^{S\bar{K}}_{\bar{a}\bar{b}}\right)~,
\end{equation}
and with ${B^\dagger}^{SK}$ defined as in Eq.~(\ref{2DM_op_transin}). Because of this $p\mathcal{K}$-ensemble definition and the fact that parity symmetry implies that,
\begin{equation}
^{\mathcal{K}}\Gamma^{SK}_{ab;cd} = ~^{-\mathcal{K}}\Gamma^{S\bar{K}}_{\bar{a}\bar{b};\bar{c}\bar{d}}~,
\end{equation}
one sees that the 2DM becomes diagonal in two-particle parity.
As was the case for the 1DM, the $\tilde{K} = 0$ and $\pi$ are mapped on themselves, but because the single-particle momenta $a,b$ change, both positive and negative parity combinations can now be formed. Let us take a look at the different possibilities:
\paragraph{$\mathbf{{\tilde{K}} = 0}$:}
for $\tilde{K} = 0$ the single-particle indices in Eq.~(\ref{tp_ti_par}) have to satisfy:
\begin{equation}
(a+b)\%2\pi = 0\qquad\text{or}\qquad a = \bar{b}~.
\end{equation}
This means that $\tilde{K} = 0$ states can be written as:
\begin{equation}
\ket{a\bar{a};S0^\pi} = {}^{\Gamma}N_{a\bar{a}}^0\left(\ket{a\bar{a};S0} + \pi\ket{\bar{a}a;S0}\right)~,
\label{tp_ti_par_0}
\end{equation}
in which the second term is equal to the first but with exchanged single-particle indices. From previous discussions we know that the symmetry of the two-particle state under the exchange of the single-particle indices depends on the two-particle spin, {\it i.e.}
\begin{equation}
\ket{a\bar{a};S0} = (-1)^S\ket{\bar{a}a;S0}~.
\end{equation}
One can see from Eq.~(\ref{tp_ti_par_0}) that for $S= 0$ only the positive parity states, and for $S=1$ only negative parity states remain. The norm is given ${}^{\Gamma}N^0_{a\bar{a}} = \frac{1}{2}$. In the $\tilde{K}=0$ case, the definition of the parity-symmetric 2DM as a function of the regular translationally invariant 2DM then reduces to:
\begin{equation}
\Gamma^{S0^\pi}_{ab;cd} = \delta_{\pi(-1)^S}\Gamma^{S0}_{ab;cd}~.
\label{2DM_ti_par_0}
\end{equation}
\paragraph{$\mathbf{0<{\tilde{K}} <\boldsymbol\pi}$:}
for $0<\tilde{K} <\pi$ the first and second term in Eq.~(\ref{tp_ti_par}) consist of different single-particle indices $a\neq \bar{b}$, implying that both positive and negative parity combinations can be constructed, with norm ${}^{\Gamma}N_{ab}^K = \frac{1}{\sqrt{2}}$. As shown for the 1DM, the $p\mathcal{K}$ ensemble makes the 2DM diagonal in, and independent of, parity. Hence every block is twofold degenerate. Since $K\neq\bar{K}$ there are only two terms remaining in the definition of the parity-symmetric 2DM:
\begin{equation}
\Gamma^{S\tilde{K}^\pi}_{ab;cd} = \frac{1}{2}\left(\Gamma^{SK}_{ab;cd}+\Gamma^{S\bar{K}}_{\bar{a}\bar{b};\bar{c}\bar{d}}\right)~.
\label{2DM_ti_par_K}
\end{equation}
\paragraph{$\mathbf{{\tilde{K}} = \boldsymbol\pi}$:}
Finally, for this block $K$ again equals $\bar{K}$. In this case there is always one state that is mapped on itself, and for which only a positive parity combination can be formed, {\it i.e.} $\ket{0\pi;S\pi^+}$, with norm ${}^{\Gamma}N_{0\pi}^\pi = \frac{1}{2}$. For all the other states in this block both positive and negative parity combinations can be formed, with norm ${}^{\Gamma}N_{ab}^\pi=\frac{1}{\sqrt{2}}$. Because of the $p\mathcal{K}$ ensemble, the 2DM falls apart in a positive and negative parity block, and since $K=\bar{K}$, four terms remain in the definition of the 2DM:
\begin{equation}
\Gamma^{S\pi^\pi}_{ab;cd} = {}^{\Gamma}N_{ab}^\pi~{}^{\Gamma}N_{cd}^\pi\left(\Gamma^{S\pi}_{ab;cd} + \Gamma^{S\pi}_{\bar{a}\bar{b};\bar{c}\bar{d}} + \pi\left[\Gamma^{S\pi}_{ab;\bar{c}\bar{d}}+\Gamma^{S\pi}_{\bar{a}\bar{b};cd}\right]\right)~.
\label{2DM_ti_par_pi}
\end{equation}
We observe from Eq.~(\ref{2DM_ti_par_pi}) that the original $K=\pi$ block reduces to a positive and negative parity block, for both the $S=0$ and $S=1$ part. Also note that there is \emph{no} degeneracy between the positive and negative parity block!

Similar considerations hold for the matrix constraints, as an example the explicit case of the $\mathcal{G}$ condition is given.

The parity-symmetric form of a particle-hole state is defined as:
\begin{equation}
\ket{ab;S\tilde{K}^\pi} =~^\mathcal{G}N^K_{ab} \left(\left[a^\dagger_a \otimes \tilde{a}_b\right]^{SK} + \left[a^\dagger_{\bar{a}} \otimes \tilde{a}_{\bar{b}}\right]^{S\bar{K}}\right)\ket{0}~,
\end{equation}
in which the hole operator $\tilde{a}_{k\sigma}$ is defined as in Eq.~(\ref{good_sto_mom}). Using this parity-symmetric particle-hole operator the $\mathcal{G}$ map is defined in a $p\mathcal{K}$ ensemble, which once again renders the matrix diagonal in particle-hole parity. The particle-hole states can be divided into two classes, on the one hand $\tilde{K}=0\text{ or }\pi$, and on the other hand those states which are mapped on a different momentum. For simplicity we first consider this last class.
\paragraph{$\mathbf{0 < \tilde{K} < \boldsymbol\pi}$:} in this case one can construct both positive and negative parity combinations, with norm $^\mathcal{G}N_{ab}^K = \frac{1}{\sqrt{2}}$. The resulting $\mathcal{G}$ matrix contains only two terms, because of momentum conservion, and is independent of particle-hole parity, so every block is twofold degenerate:
\begin{equation}
\mathcal{G}(\Gamma)^{S\tilde{K}^\pi}_{ab;cd} = \frac{1}{2}\left[\mathcal{G}(\Gamma)^{SK}_{ab;cd}+\mathcal{G}(\Gamma)^{S\bar{K}}_{\bar{a}\bar{b};\bar{c}\bar{d}}\right]~.
\end{equation}
This implies the following expression of the $\mathcal{G}$ map as a function of the parity-symmetric 2DM:
\begin{align}
\mathcal{G}(\Gamma)^{S\tilde{K}^\pi}_{ab;cd} =& \delta_{ac}\delta_{bd}\rho_{\tilde{a}}-\frac{\sqrt{(1 + \delta_{a\bar{d}})(1 + \delta_{c\bar{b}})}}{4~{}^{\Gamma}N^{K'}_{a\bar{d}}~{}^{\Gamma}N^{K'}_{c\bar{b}}}\sum_{S'}[S']^2
\left\{
\begin{matrix}
\frac{1}{2}&\frac{1}{2}&S\\
\frac{1}{2}&\frac{1}{2}&S'
\end{matrix}
\right\}
\sum_{\pi'}
\Gamma^{S'\tilde{K}'^{\pi'}}_{a\bar{d};c\bar{b}}~.
\end{align}
\paragraph{$\mathbf{\tilde{K} = 0}$ and $\mathbf{\tilde{K} = \boldsymbol\pi}$:} both the $\tilde{K} = 0$ and $\tilde{K} = \pi$ blocks are mapped on themselves. For $\tilde{K} = 0$ the action of the parity operator is again to exchange the single-particle momenta, but in contrast with the two-particle case, there is no symmetry between the particle and the hole index. As a consequence positive and negative parity combinations for both $\tilde{K} = 0$ and $\tilde{K} = \pi$ can be constructed, with norms $^\mathcal{G}N_{ab}^K=\frac{1}{\sqrt{2}}$. There are a few exceptions however: for $\tilde{K} = 0$, the states with $a = b = 0$ and $a = b = \pi$, and for $\tilde{K} = \pi$ the states with $a = 0$, $b = \pi$ and $a=\pi,b= 0$, are mapped on themselves and only occur in the positive parity block, with norm $^{\mathcal{G}}N_{ab}^{K} = \frac{1}{2}$. The general form of the parity-symmetric $\mathcal{G}$ map when ${K} ={\bar{K}}$, as a function of the regular translationally invariant $\mathcal{G}$ is:
\begin{align}
\mathcal{G}(\Gamma)^{S\tilde{K}^\pi}_{ab;cd} =& ~^{\mathcal{G}}N^K_{ab}~^{\mathcal{G}}N^K_{cd}\left[
\mathcal{G}(\Gamma)^{SK}_{ab;cd}
+\mathcal{G}(\Gamma)^{SK}_{\bar{a}\bar{b};\bar{c}\bar{d}}
+\pi\left(\mathcal{G}(\Gamma)^{SK}_{ab;\bar{c}\bar{d}}
+\mathcal{G}(\Gamma)^{SK}_{\bar{a}\bar{b};cd}
\right)
\right]~.
\end{align}
In this case the expression of $\mathcal{G}$ as a function of the 2DM is a bit more complicated:
\begin{align}
\nonumber\mathcal{G}(\Gamma)^{S\tilde{K}^\pi}_{ab;cd} =& \delta_{ac}\delta_{bd}\rho_{\tilde{a}}
-~^\mathcal{G}N^K_{ab} ~^\mathcal{G}N^K_{cd}
\sum_{S'}[S']^2
\left\{
\begin{matrix}
\frac{1}{2}&\frac{1}{2}&S\\
\frac{1}{2}&\frac{1}{2}&S'
\end{matrix}
\right\}
\left[
\frac{\sqrt{(1 + \delta_{a\bar{d}})(1 + \delta_{c\bar{b}})}}{4~{}^{\Gamma}N^{K'}_{a\bar{d}}{}^{\Gamma}N^{K'}_{c\bar{b}}}
\sum_{\pi'}
\Gamma^{S'\tilde{K}'^{\pi'}}_{a\bar{d};c\bar{b}}\right.\\
&\left. \qquad\qquad\qquad\qquad\qquad\qquad+ \pi\frac{\sqrt{(1 + \delta_{a{d}})(1 + \delta_{c{b}})}}{4~{}^{\Gamma}N^{K''}_{\bar{a}\bar{d}}~{}^{\Gamma}N^{K''}_{c{b}}}
\sum_{\pi'}
\Gamma^{S'\tilde{K}''^{\pi''}}_{\bar{a}\bar{d};c{b}}\right]~.
\end{align}

\section{\label{computational}Computational performance}
%drie methodes vergelijken, toon dat de schaling het beste is voor boundary point
%TODO misschien nog zeggen dat voor de T condities dat niet zo is.
\begin{figure}
\centering
\includegraphics[scale=0.8]{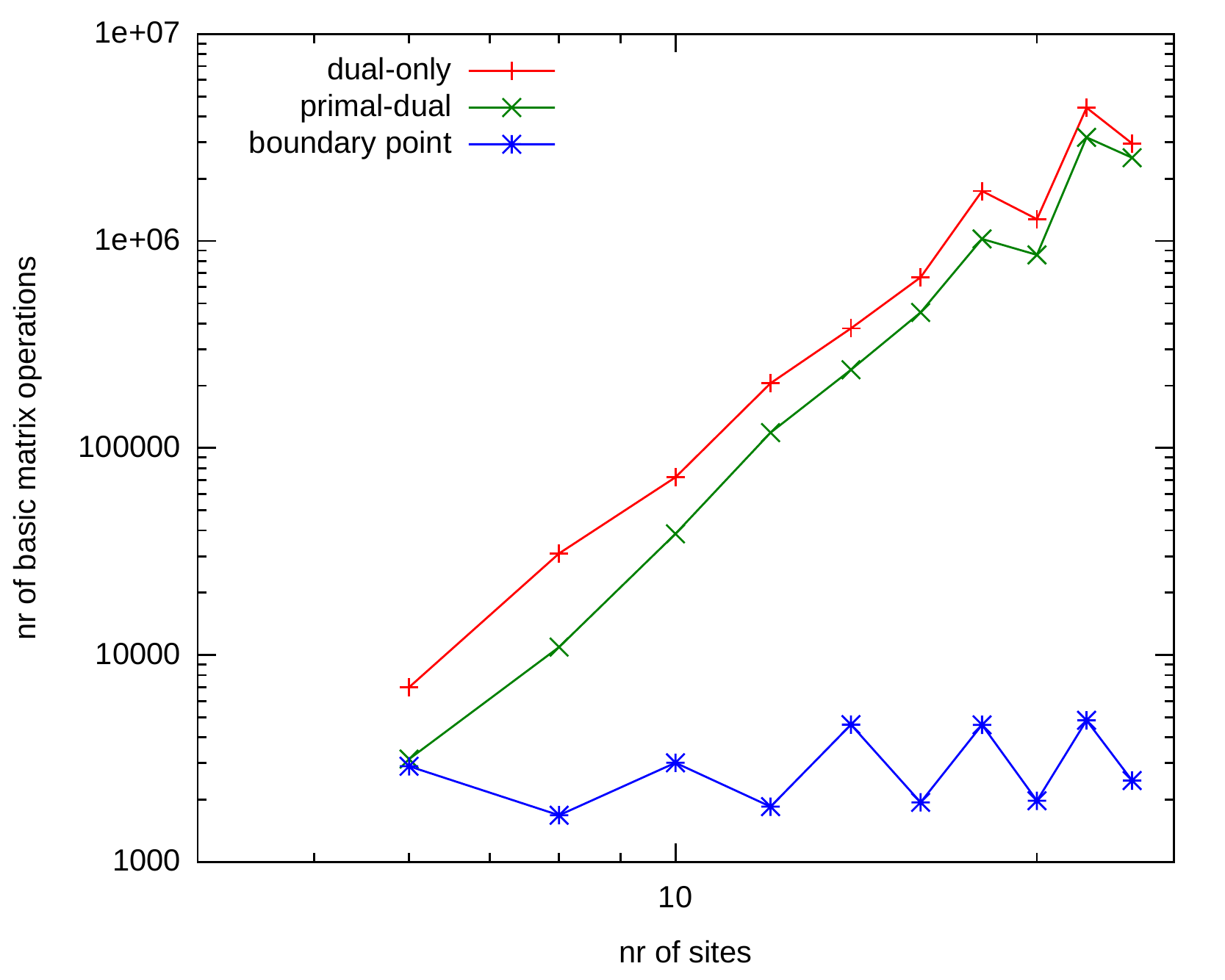}
\caption{\label{scaling} The number of basic matrix operations needed to converge for different lattice sizes of the one-dimensional Hubbard model, using $\mathcal{IQG}$ conditions and $U = 1$.}
\end{figure}
In v2DM we want to optimize the energy by varying a matrix, the 2DM, under the constraints that it has the right particle number, and that some linear matrix maps of the 2DM are positive semidefinite, {\it i.e.}
\begin{eqnarray}
\label{v2DM}E^N_{\text{SDP}}\left(H_\nu\right) &=& \min_{\Gamma} \mathrm{Tr}~\left[\Gamma H^{(2)}_\nu\right]\\
\text{u.c.t.}&&\left\{
   \begin{array}{l}
   \mathrm{Tr}~\Gamma = \frac{N(N-1)}{2}~,\\
   \mathcal{L}_j\left(\Gamma\right) \succeq 0~.
   \end{array}
   \right.
   \label{2DM_constraints}
\end{eqnarray}
Here the $\mathcal{L}_j$ are a collection of matrix non-negativity constraints to be applied. It is well known that this problem can be reformulated as a semidefinite program. There is a vast literature on this subject and many different algorithms exist to solve this type of problem. Because of the large amount of symmetry present in the Hubbard model, there is a huge block diagonalization of the 2DM and the matrix constraints. This leads to a significant reduction of the computational cost of a basic matrix computation. This feature has allowed us to go up to large lattice sizes and compare the computational scaling of different algorithms. We have implemented three different semidefinite programming algorithms. Two are so-called interior point methods (a dual-only potential reduction algorithm, \cite{atomic} and a primal-dual interior point algorithm \cite{primal_dual}), where the 2DM is optimized from within the $N$-representable region. In the third one (a boundary point method \cite{maz_bp}) the 2DM is not required to be $N$-representable during the optimization. In this Section we compare the computational scaling of these methods for the one-dimensional Hubbard model with $U = 1$ and using $\mathcal{IQG}$ conditions. For details about the implementation of the different algorithms, see the cited references.

All the methods have the same basic computational scaling behaviour, being $O(M^6)$ (with $M$ the dimension of single-particle Hilbert space) required for multiplying, inverting or diagonalizing a matrix. In Figure~\ref{scaling} the number of these operations needed to converge to the optimum is plotted as a function of lattice size. The interior point methods both have to solve a linear system of size $M^4$, so it is not surprising that the scaling, on top of the basic matrix computations, of these methods is $M^4$. More surprising is that there seems to be no, or a very limited, scaling for the boundary point method. The number of iterations required remains around 3000, irrespective of the size of the system. It must be stressed that this is a result limited to the one-dimensional hubbard model, and cannot be extrapolated to other systems, like molecules, where the convergence properties of the boundary point method can be completely different. One reason for the succes of the boundary point method applied to the Hubbard model is the amount of symmetry present in the system. The boundary point method is designed for problems with a huge amount of dual variables or primal constraints. For most physical systems the dimensions of the matrices involved are already unfeasibly large before the boundary point method would becomes advantageous. The Hubbard model, however, contains many symmetries, implying that the matrix dimensions are considerably reduced, and the number of dual variables can get very large before the matrix computations involved become unfeasible. In this case, the domain where the boundary point method is advantageous is actually reached.
\section{Results}
In this Section we present and discuss the results of v2DM calculations, taking advantage of all the symmetries, on a 50-site lattice with the $\mathcal{IQG}$ conditions, and on a 20-site lattice with the $\mathcal{IQG}\mathcal{T}_1\mathcal{T}_2$ ($\mathcal{IQGT}$) conditions. The Hubbard model has been studied before using the v2DM method, see {\it e.g.} \cite{maz_hub,shenvi,nakata_last,gutz_sdp}, but only the energy was considered, and this for relatively small lattice sizes (up to $L = 14$). In this paper we study different filling factors, and extract various properties like the ground-state energy and two-particle correlation functions in order to assess the quality of the variationally obtained 2DM. 
The v2DM results discussed in this Section were all obtained using the primal-dual predictor corrector semidefinite programming algorithm \cite{primal_dual}. Although the one-dimensional Hubbard model can be solved exactly using the Bethe ansatz \cite{bethe,liebwu,essler,1D_hub}, it is hard to extract information about the solution for finite systems. For the calculations on a 20-site lattice, we compare the data with the quasi-exact results obtained through a variational Matrix Product State (MPS) algorithm \cite{schollwock,verstraete,chan}, written by co-worker Sebastian Wouters \cite{sebastian}. For the 50-site lattice, however, this is no longer computationally feasible. At half filling a simplification in the Bethe-ansatz equations occurs, which allows to calculate the ground-state energy of finite systems by solving a set of non-linear equations (Lieb-Wu)\cite{liebwu_2}. At other fillings no data is available for comparison.
\subsection{Ground-state energy}
\begin{figure}
\centering
$
\begin{array}{c}
\includegraphics[scale=0.7]{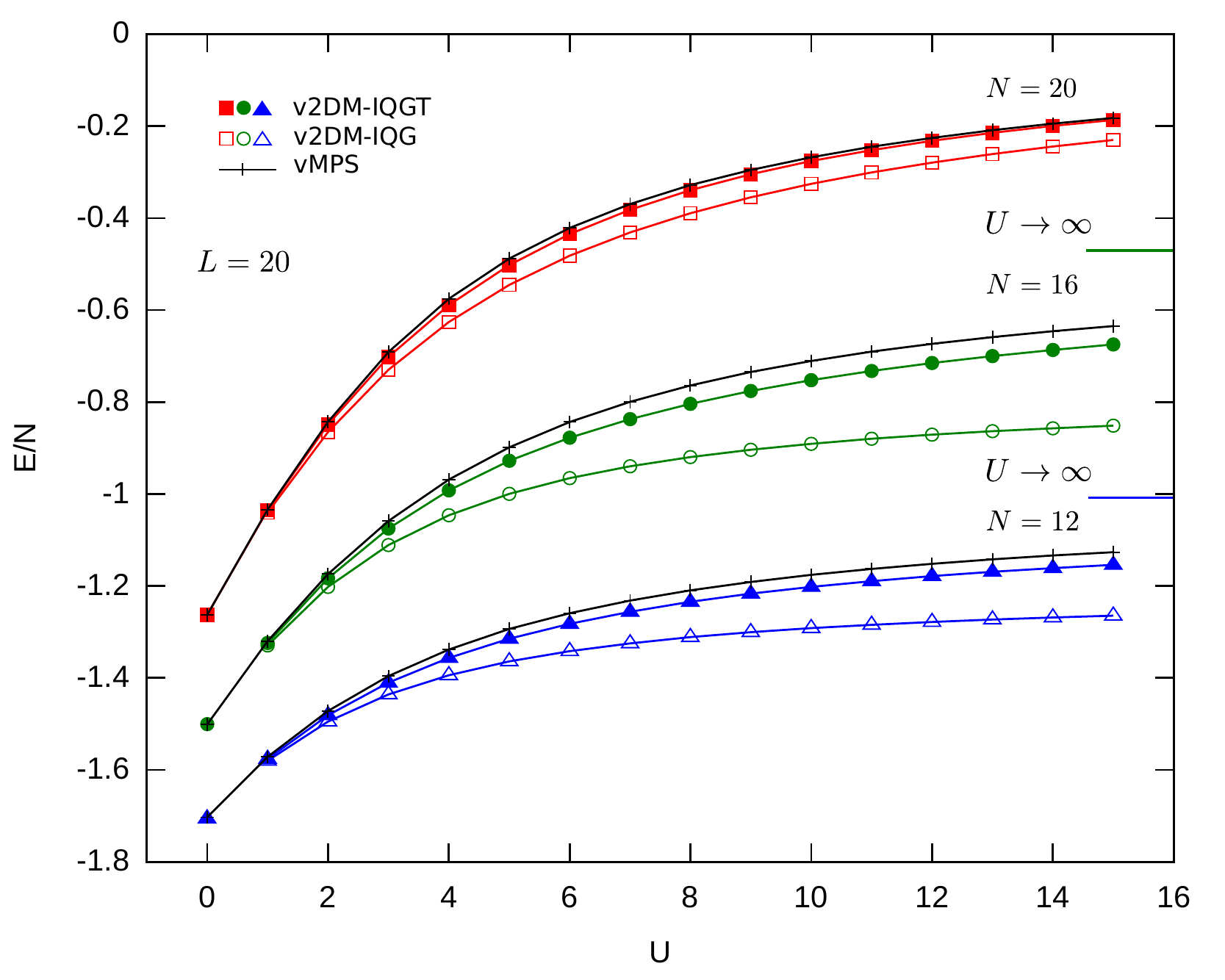}\\
\includegraphics[scale=0.7]{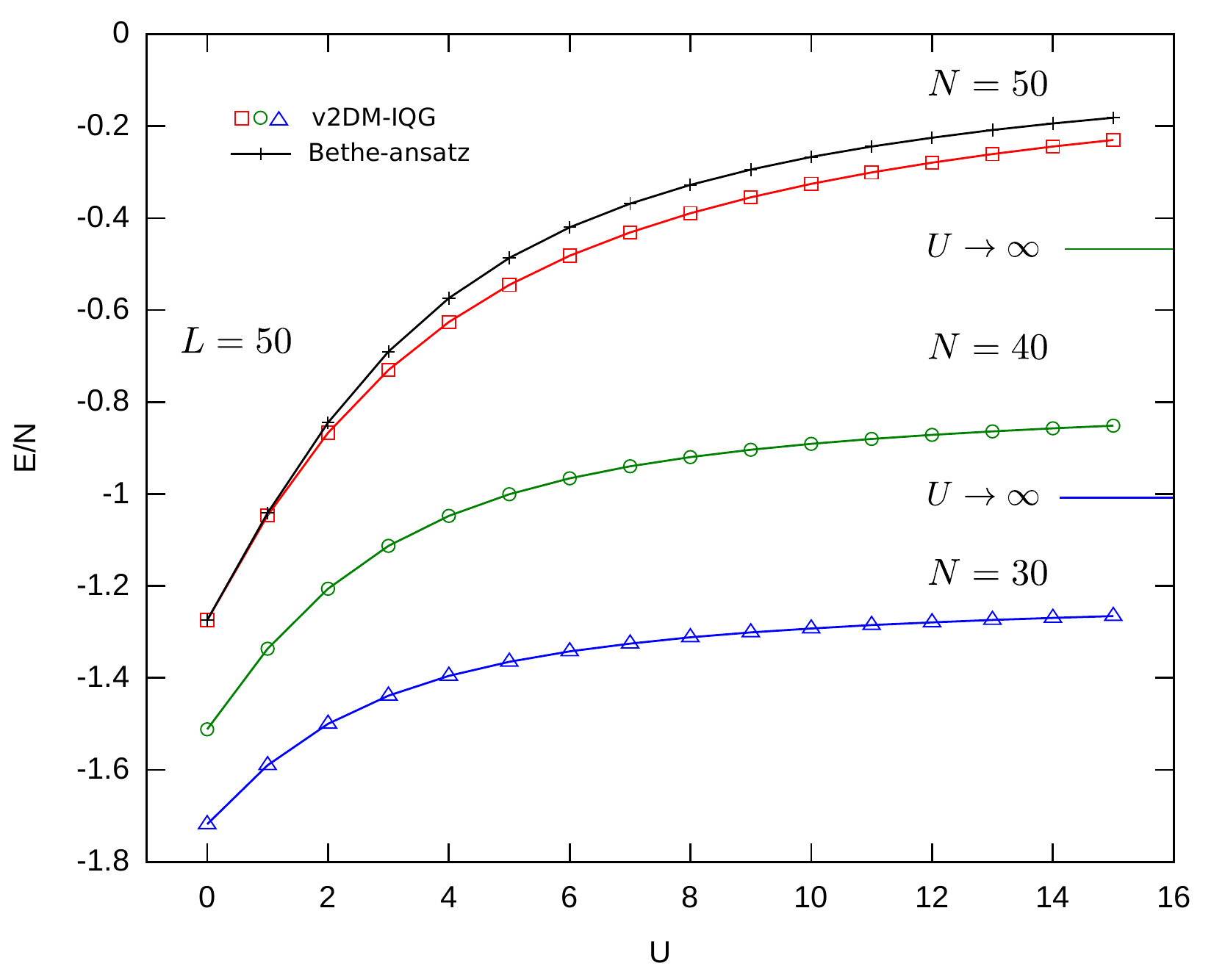}
\end{array}
$
\caption{\label{energy_1dhub}Ground-state energy per particle as a function of on-site repulsion $U$ of the Hubbard model for a 20-site (top) and 50-site (bottom) lattice at half-, $\frac{4}{10}$  and $\frac{3}{10}$ filling. For the 20-site lattice a comparison is made between v2DM using the $\mathcal{IQG}$ and $\mathcal{IQGT}$ conditions, and a quasi-exact result using MPS. For the 50-site lattice only $\mathcal{IQG}$ conditions are feasible, and these have been compared to the exact (Bethe-ansatz) result for half filling.}
\end{figure}
In Fig.~\ref{energy_1dhub} the ground-state energy per particle of the one-dimensional Hubbard model is plotted as a function of the on-site repulsion $U$ (the hopping parameter $t$ will always be taken equal to unity). In the top figure the v2DM results for the 20-site lattice are shown for three different fillings, 12 particles ($\frac{3}{10}$), 16 particles ($\frac{4}{10}$) and half filling. These were calculated using both the $\mathcal{IQG}$ and the $\mathcal{IQGT}$ conditions, and are compared to the quasi-exact variational MPS results. In the bottom figure the v2DM results for the 50-site lattice are shown for the same fillings ({\it i.e.} 30 particles ($\frac{3}{10}$), 40 particles ($\frac{4}{10}$) and half filling). For the 50-site lattice it was only possible to perform the calculations using the $\mathcal{IQG}$ conditions, and compare to the exact solution obtained by solving the Lieb-Wu equations for the half-filled lattice \cite{liebwu_2}. 

\begin{figure}
\centering
\includegraphics[scale=0.7]{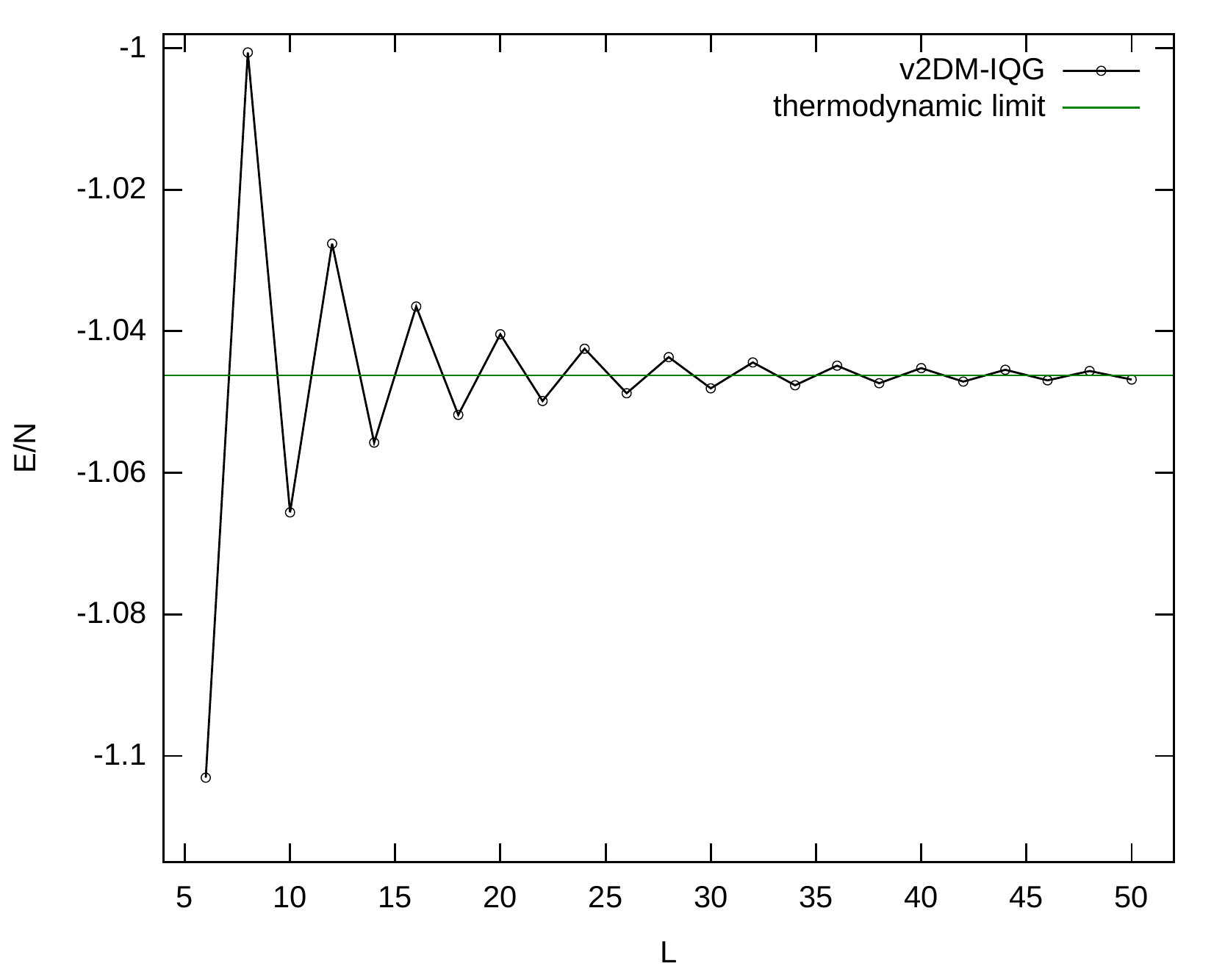}
\caption{\label{enconv}Ground-state energy per particle as a function of the lattice size $L$, indicating how fast the finite size results converge to the thermodynamic limit for half filling.}
\end{figure}

One interesting thing to notice is that the $\mathcal{IQG}$ energy per particle for the 20-site lattice and the 50-site lattice, at the same filling, are very similar. This is due to the periodic boundary conditions which make the results converge quite rapidly for increasing lattice size $L$, implying that one can already extract relevant results for the thermodynamic limit by studying relatively small lattices. This fast convergence can be clearly seen in Fig.~\ref{enconv}, where we plotted the energy per particle of a Hubbard model with $U=1$ at half filling, as a function of the lattice size $L$. This result seems to indicate that the method is more or less size extensive for the Hubbard model, which is surprising since in general v2DM is not size extensize \cite{nakata_se,helen_1}.

Another thing to remark in Fig.~\ref{energy_1dhub} is that for the 20-site lattice, the difference between $\mathcal{IQG}$ and $\mathcal{IQGT}$ is rather small for the half-filled lattice, but larger for the other fillings, and that the difference gets larger when $U$ increases. For the 50-site lattice we see that the $\mathcal{IQG}$ result agrees nicely with the solution of the Lieb-Wu equations. For the other fillings no reference data are available. There are, however, two limits of the model that are exactly solvable. The first limit is the rather trivial case of no interaction, {\it i.e.} $U=0$, for which the solution has already been given in Eq.~(\ref{sp_hubbard}). The Hamiltonian reduces to a single-particle operator, which means this limit is already described correctly by including the $\mathcal{I}$ and $\mathcal{Q}$ conditions alone. The other exactly solvable limit is when $U\rightarrow+\infty$. In this limit the physics of the model decouples into two independent parts, one describing the spin of the system, and the other the movement of the particles (this is called spin-charge separation \cite{ogata}). This decoupling shows up in the Bethe-ansatz wave function: the charge degrees of freedom are described by a Slater determinant of spinless fermions, whereas the spin degrees of freedom become equivalent to a spin-$\frac{1}{2}$ Heisenberg model. The single-particle energy spectrum changes slightly compared to Eq.~(\ref{sp_hubbard}) because the boundary conditions for spinless fermions are periodic/antiperiodic if $N$ is even/odd \cite{ogata,krivnov}:
\begin{equation}
\epsilon_k = -2t\cos{k}\qquad\text{where}\qquad
\left\{
\begin{matrix}
k = \frac{2\pi n}{L}&\qquad\text{if}\qquad N\%2=0\\
k = \frac{(2n+1)\pi}{L}&\qquad\text{if}\qquad N\%2=1
\end{matrix}
\right.~.
\label{1dhub_sclim}
\end{equation}
When the lattice is half-filled all the single-particle states are occupied, and the total energy sums up to zero, which is correctly described by the $\mathcal{IQG}$ results in Fig.~\ref{energy_1dhub}. Away from half filling, however, the energy has a finite limit which can be calculated using Eq.~(\ref{1dhub_sclim}). From the figure we can see that the $\mathcal{IQG}$ conditions do not suffice to correctly describe the large-$U$ limit. Only when the $\mathcal{T}$ conditions are added, the results converge to the right limit. Calculations at very large values of $U$ have been performed that confirm this statement, and these results are shown in Table~\ref{hub_largeU}.
\begin{table}
\centering
\caption{\label{hub_largeU}Energy per site of the v2DM calculations away from half filling at large values of $U$, compared to the MPS and the Bethe-ansatz results where available.}
\begin{tabular}{|c|c|ccccc|}
\hline
$L$&$N$&$U$&$\mathcal{IQG}$&$\mathcal{IQGT}$&vMPS&exact\\
\hline
\multirow{6}{*}{20}&
\multirow{3}{*}{12}&50&-1.2259&-1.0804&-1.0488&*\\
&&100&-1.2177&-1.0646&-1.03116&*\\
&&$\infty$&*&* &* &-1.0008\\
\cline{2-7}
&\multirow{3}{*}{16}&50&-0.7972&-0.5458&-0.5205&*\\
&&100&-0.7860&-0.5179&-0.49513&*\\
&&$\infty$&*&* &* &-0.4639\\
\hline
\end{tabular}
\begin{tabular}{|c|c|ccc|}
\hline
$L$&$N$&$U$&$\mathcal{IQG}$&exact\\
\hline
\multirow{6}{*}{50}&
\multirow{3}{*}{30}&50&-1.2272&*\\
&&100&-1.2191&*\\
&&$\infty$&*&-1.0008\\
\cline{2-5}
&\multirow{3}{*}{40}&50&-0.7974&*\\
&&100&-0.7862&*\\
&&$\infty$&*&-0.4671\\
\hline
\end{tabular}
\end{table}
\subsection{Correlation functions}
Two-particle correlation functions are important quantities in the analysis of lattice systems, because they usually display the physics (for instance the appearance of magnetism) present in the system . In this Section we show that in our approach, these correlation functions are easily extracted from the 2DM, and compare our results to those in \cite{sorella,ogata}.
\paragraph{Charge correlation}
\begin{figure}
\centering
$
\begin{array}{c}
\includegraphics[scale=0.7]{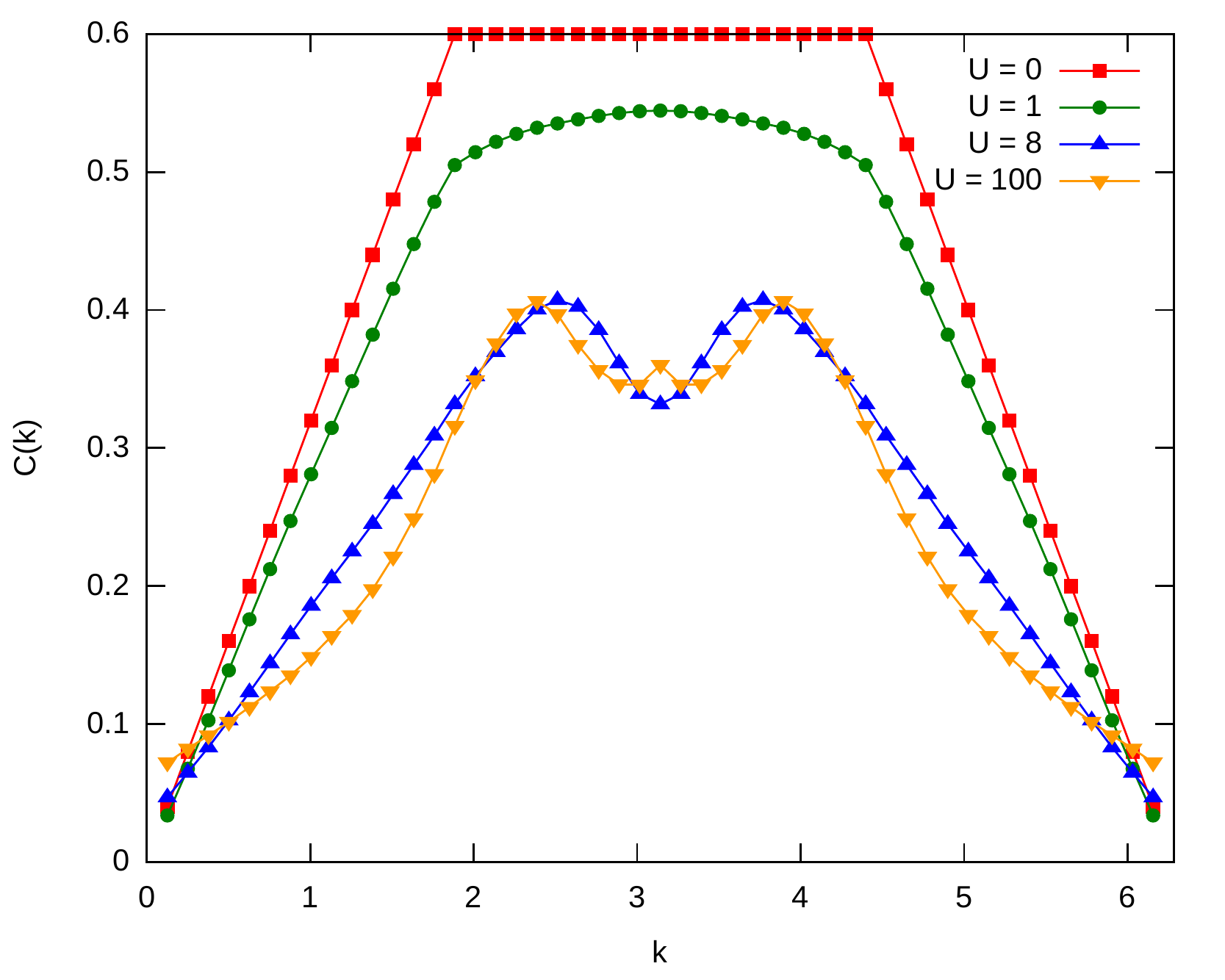}\\
\includegraphics[scale=0.7]{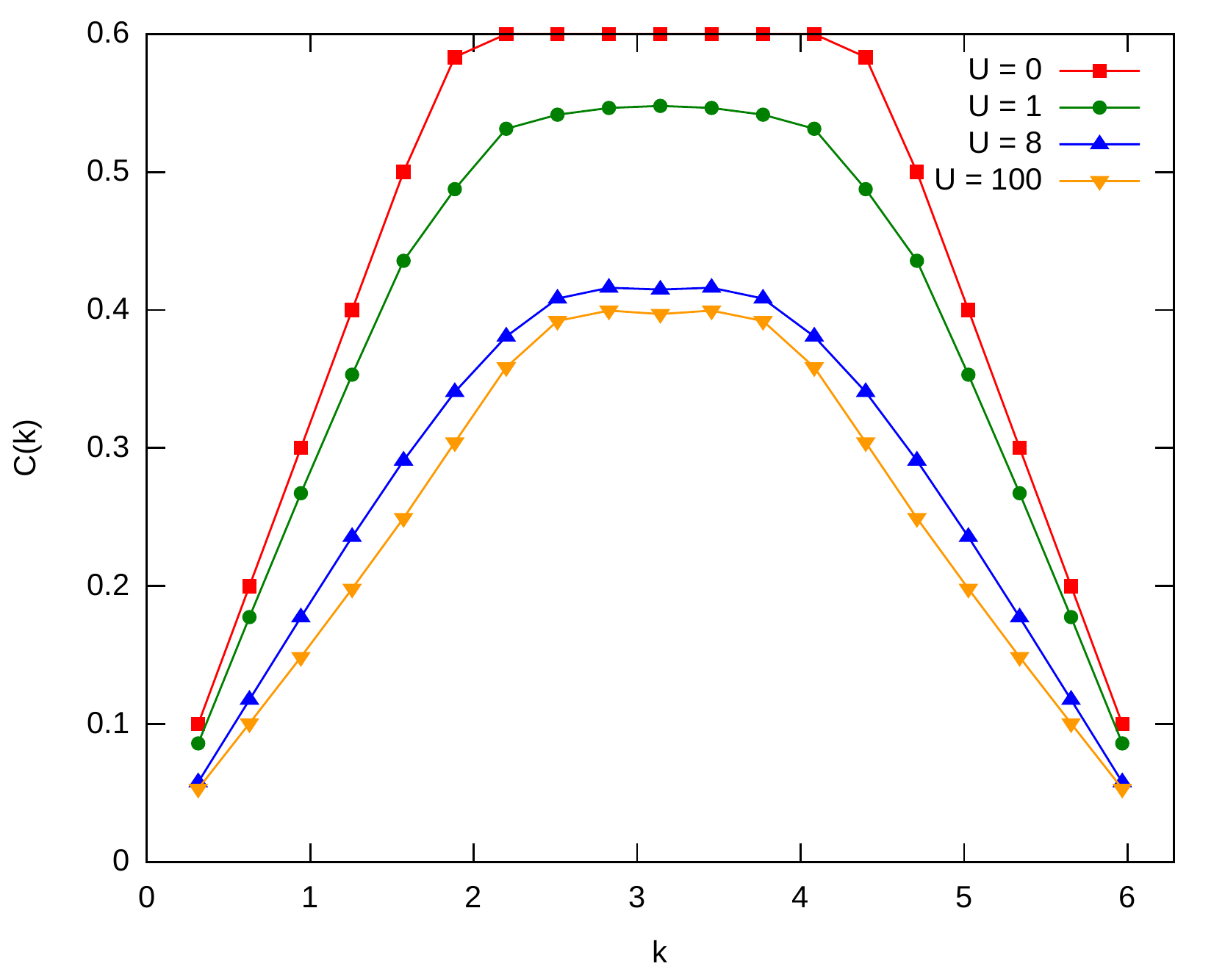}
\end{array}
$
\caption{\label{chargecorr_0.3}Two-particle charge correlation function $C(k)$, as a function of momentum, for a $\frac{3}{10}$ filled lattice and various values of on-site repulsion $U$, using $\mathcal{IQG}$ (top) and $\mathcal{IQGT}$ (bottom) conditions.}
\end{figure}
\begin{figure}
\centering
$
\begin{array}{c}
\includegraphics[scale=0.7]{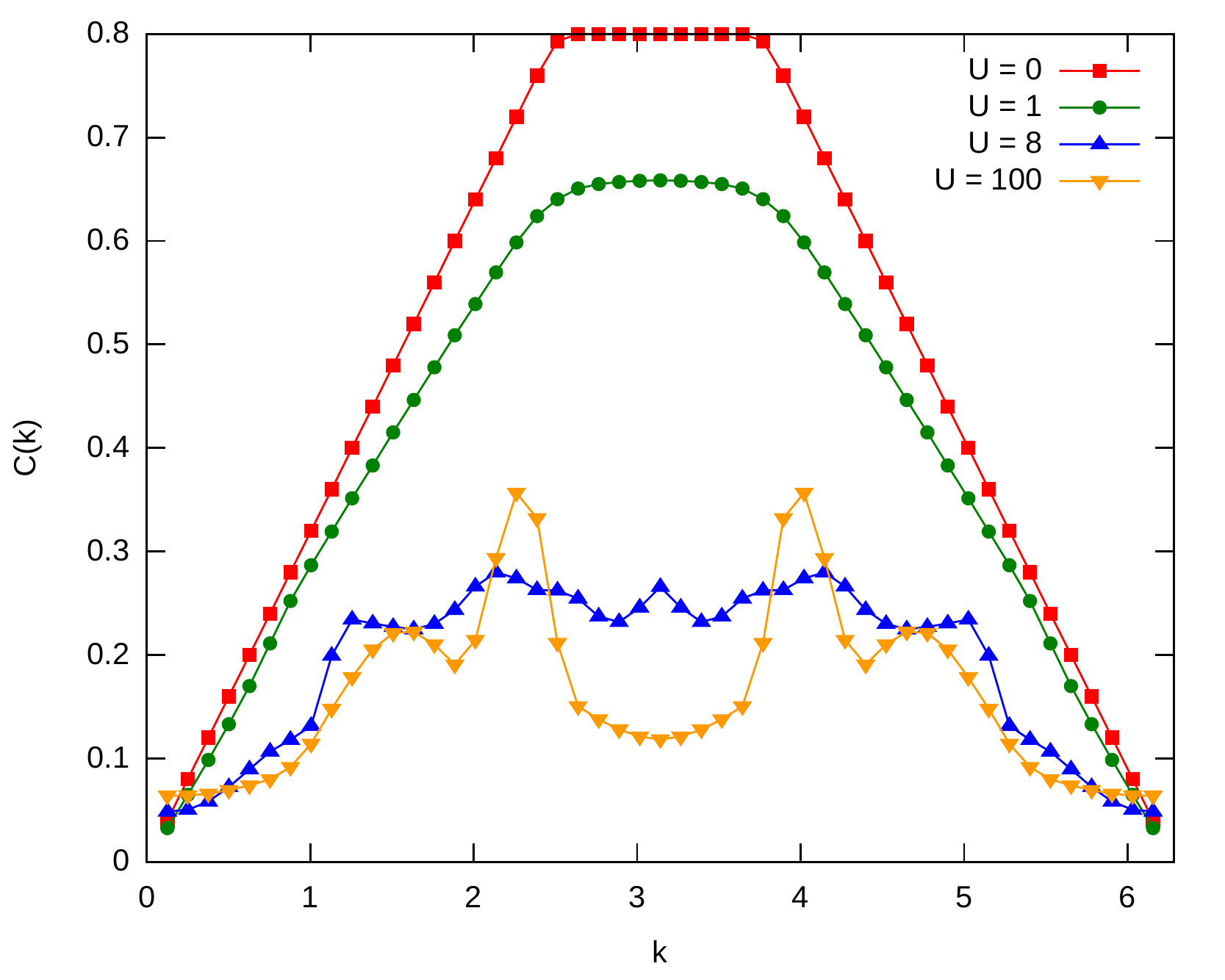}\\
\includegraphics[scale=0.7]{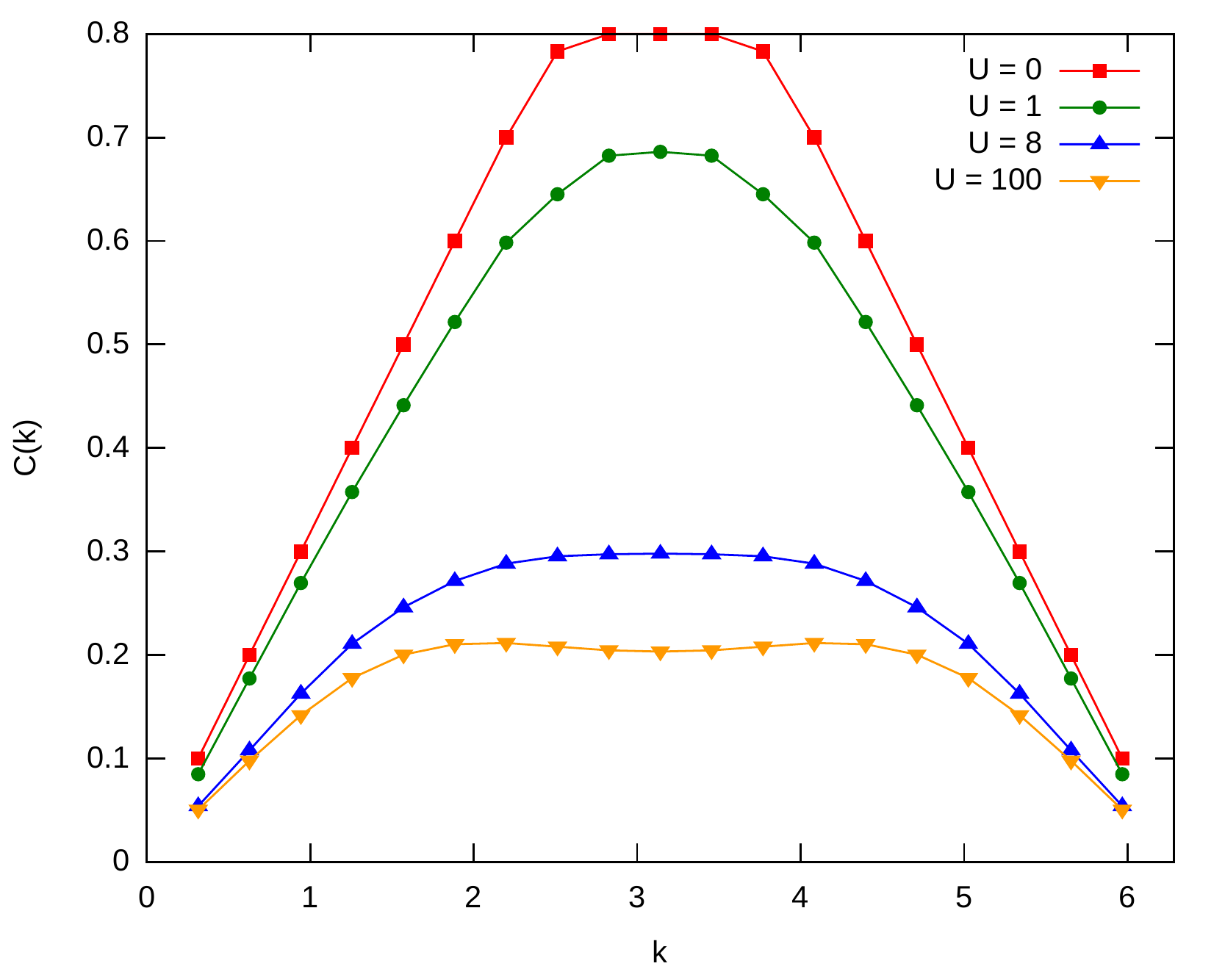}
\end{array}
$
\caption{\label{chargecorr_0.4}Two-particle charge correlation function $C(k)$, as a function of momentum, for a $\frac{4}{10}$ filled lattice and various values of on-site repulsion $U$, using $\mathcal{IQG}$ (top) and $\mathcal{IQGT}$ (bottom) conditions.}
\end{figure}
\begin{figure}
\centering
$
\begin{array}{c}
\includegraphics[scale=0.7]{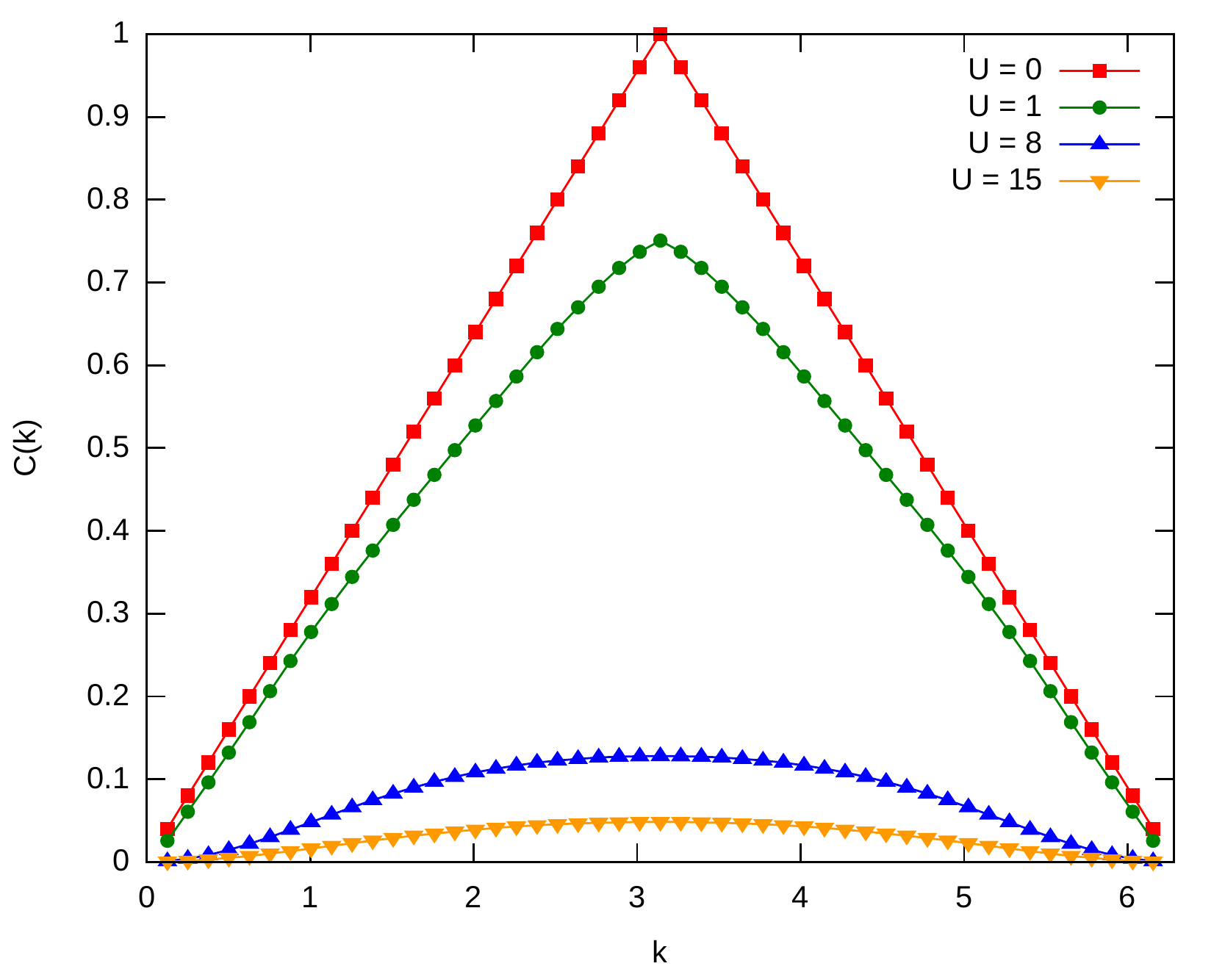}\\
\includegraphics[scale=0.7]{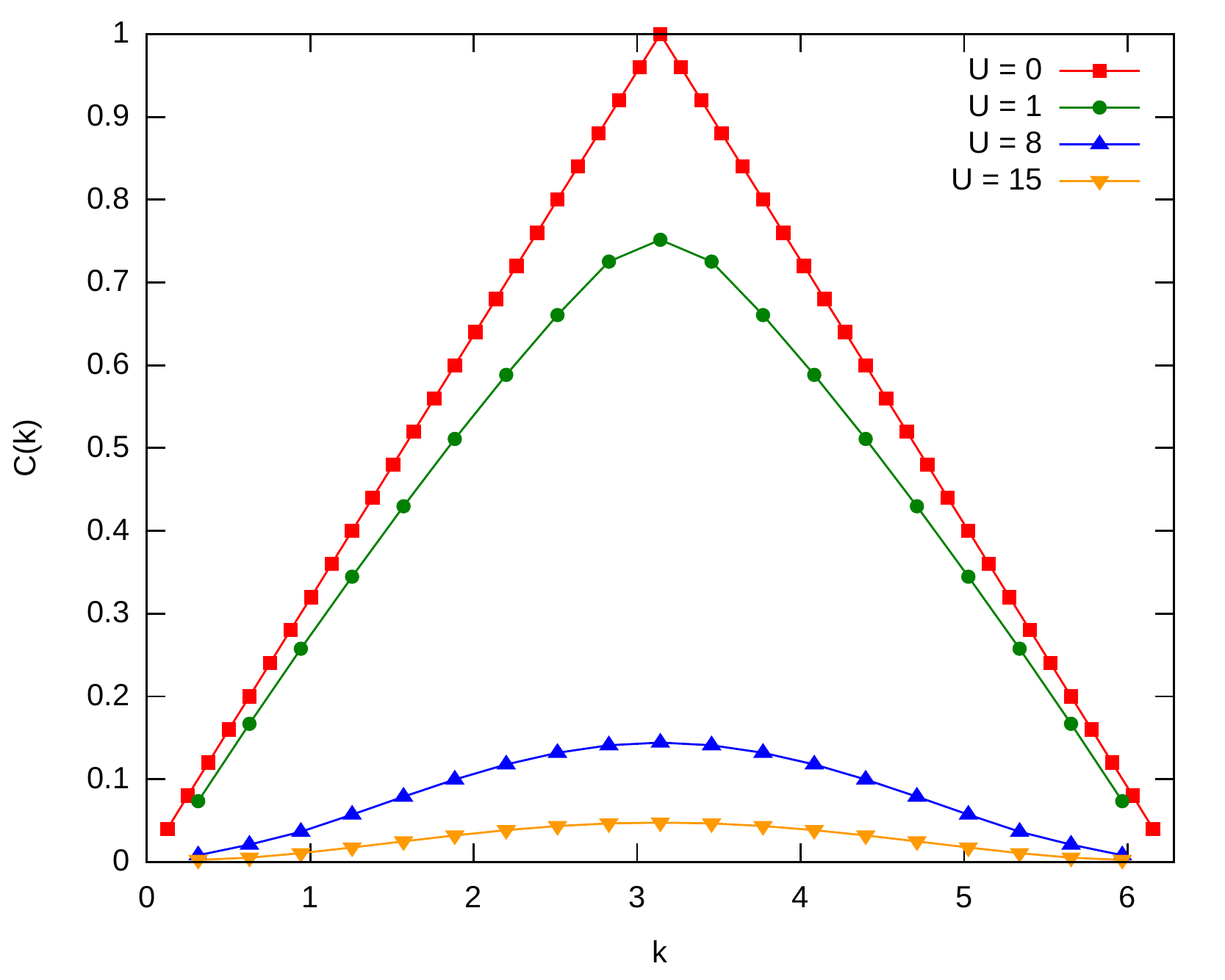}
\end{array}
$
\caption{\label{chargecorr_0.5}Two-particle charge correlation function $C(k)$, as a function of momentum, for a half-filled lattice and various values of on-site repulsion $U$, using $\mathcal{IQG}$ (top) and $\mathcal{IQGT}$ (bottom) conditions.}
\end{figure}
The two-particle charge correlation function is defined as:
\begin{equation}
C(r) = \langle \hat{n}_j\hat{n}_{j+r}\rangle = \sum_{\sigma\sigma'}\langle a^\dagger_{j\sigma}a_{j\sigma}a^\dagger_{j+r;\sigma'}a_{j+r;\sigma'} \rangle~,
\label{tpcorr}
\end{equation}
in which the notation $\langle . \rangle$ denotes the expectation value. The function is independent of the specific choice of the index $j$ because of the periodic boundary conditions. The expression in Eq.~(\ref{tpcorr}) can be written in terms of the $\mathcal{G}(\Gamma)$ matrix:
\begin{equation}
C(r) = \sum_{\sigma\sigma'} \mathcal{G}(\Gamma)_{j\sigma j\sigma;(j+r)\sigma'(j+r)\sigma'}~,
\end{equation}
and in fact only the singlet part of the $\mathcal{G}$ matrix appears:
\begin{equation}
C(r) = 2~\mathcal{G}(\Gamma)^{0}_{j j;(j+r)(j+r)}~.
\end{equation}
In translationally invariant systems one usually takes the Fourier transform of the correlation function,
\begin{equation}
C(k) = \sum_r e^{ikr} C(r) = 2\sum_{k_ak_b}\sum_{k_ck_d}\mathcal{G}(\Gamma)^{0k}_{k_ak_b;k_ck_d}~.
\end{equation}
In Figs.~\ref{chargecorr_0.3}, \ref{chargecorr_0.4} and \ref{chargecorr_0.5}, $C(k)$ has been plotted for $\frac{3}{10}$, $\frac{4}{10}$  and half filling respectively, using both $\mathcal{IQG}$ and $\mathcal{IQGT}$ conditions. Comparing the $\mathcal{IQG}$ with the $\mathcal{IQGT}$ results the same trends can be noticed as for the energy and the momentum distributions. For half filling (Fig.~\ref{chargecorr_0.5}) the $\mathcal{IQG}$ and $\mathcal{IQGT}$ results are in nice agreement. Moving away from half-filling (Figs.~\ref{chargecorr_0.3} and \ref{chargecorr_0.4}) there is only agreement for small values of $U$. For larger values of $U$ strange oscillations appear in the $\mathcal{IQG}$ results. So in this limit not only the energy, but the entire physical content of the $\mathcal{IQG}$-2DM cannot be trusted. This is once again an indication that the $\mathcal{IQG}$ conditions fail to describe the strong-correlation limit away from half-filling. The $\mathcal{IQGT}$ results compare well, both in shape and magnitude, with the results from Quantum Monte Carlo \cite{sorella}, and the Bethe-ansatz results in the strong-correlation limit \cite{ogata}.
\paragraph{Spin correlation}
\begin{figure}
\centering
$
\begin{array}{c}
\includegraphics[scale=0.7]{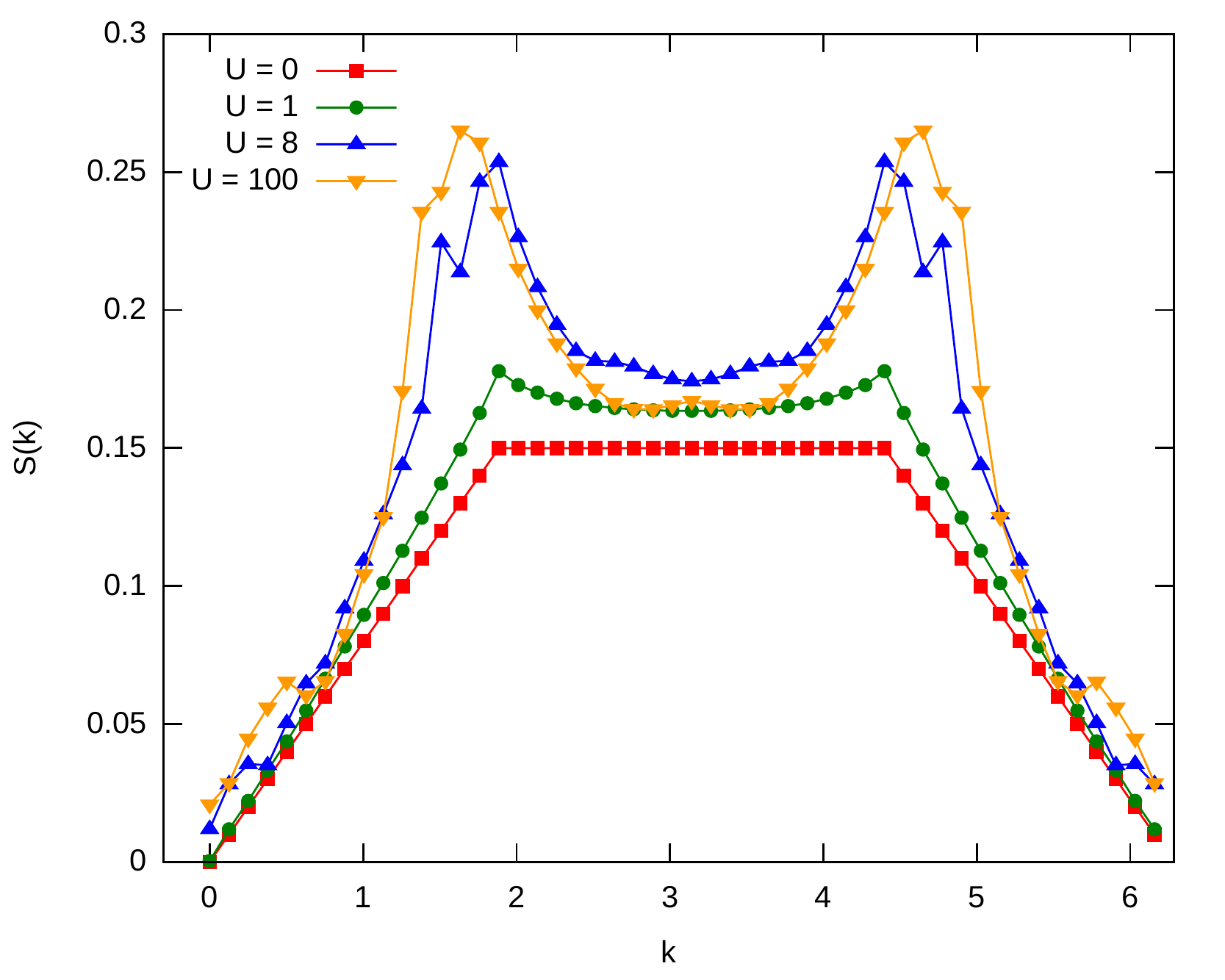}\\
\includegraphics[scale=0.7]{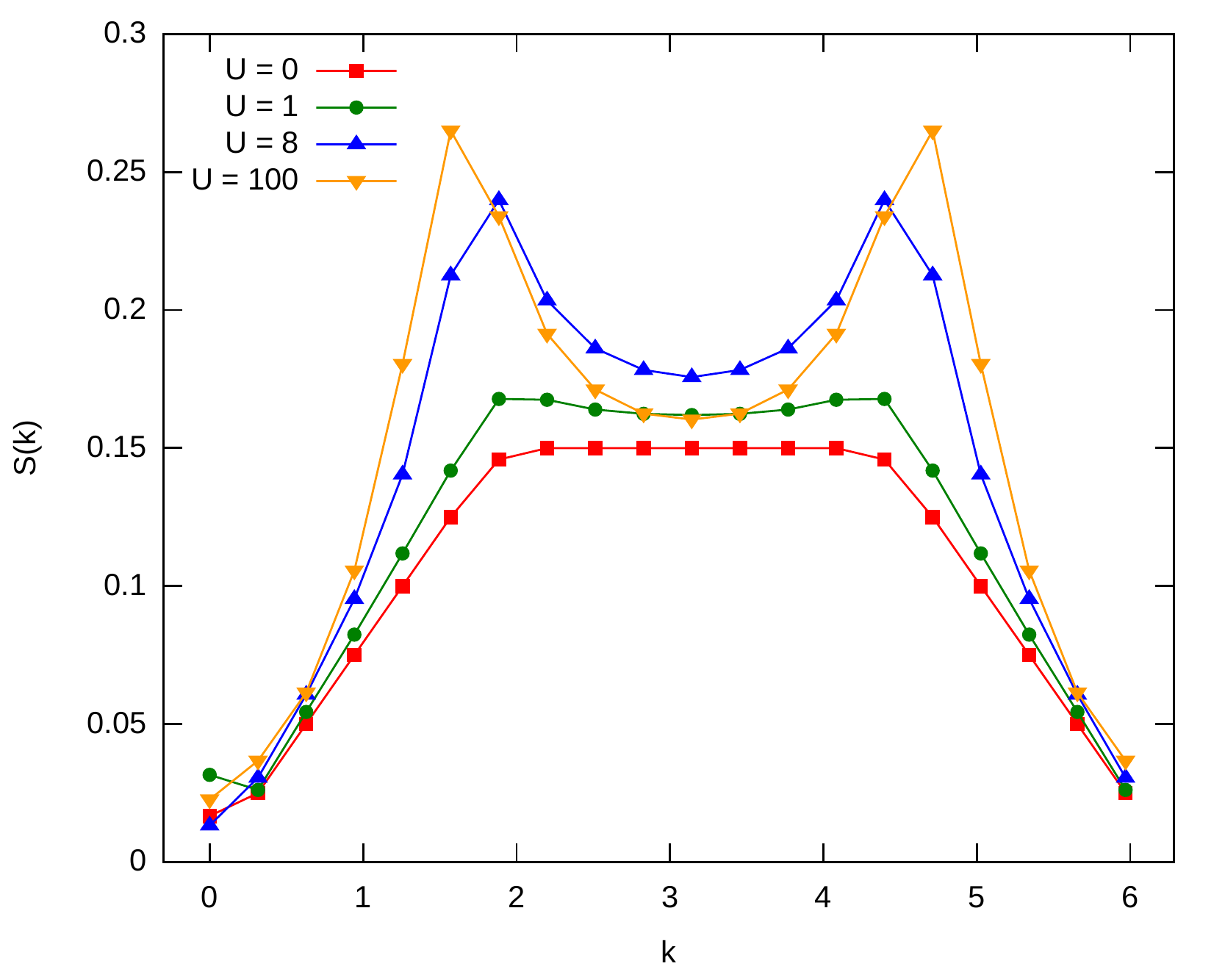}
\end{array}
$
\caption{\label{spincorr_0.3}Two-particle spin correlation function $S(k)$, as a function of momentum, for a $\frac{3}{10}$ filled lattice and various values of on-site repulsion $U$, using $\mathcal{IQG}$ (top) and $\mathcal{IQGT}$ (bottom) conditions.}
\end{figure}
\begin{figure}
\centering
$
\begin{array}{c}
\includegraphics[scale=0.7]{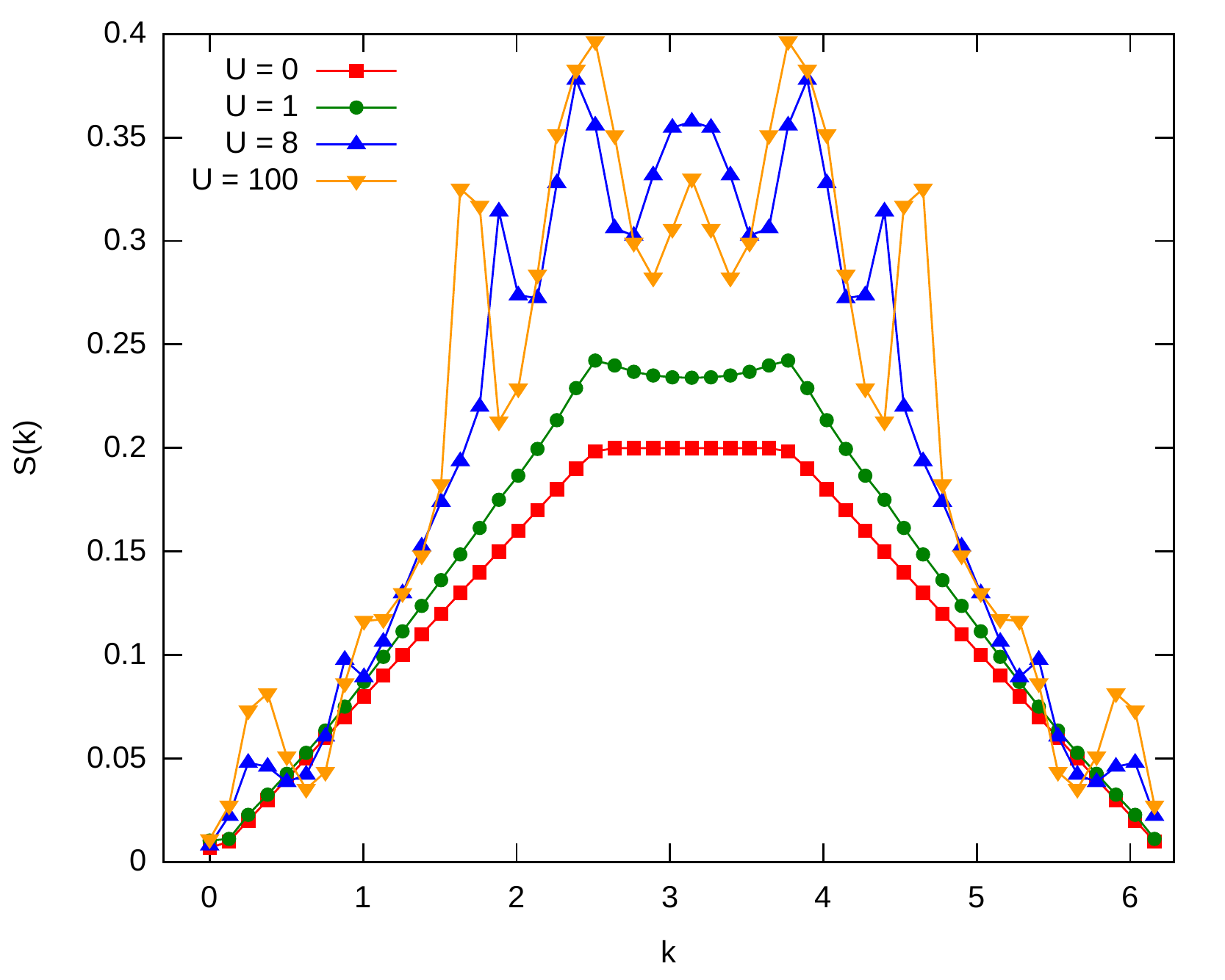}\\
\includegraphics[scale=0.7]{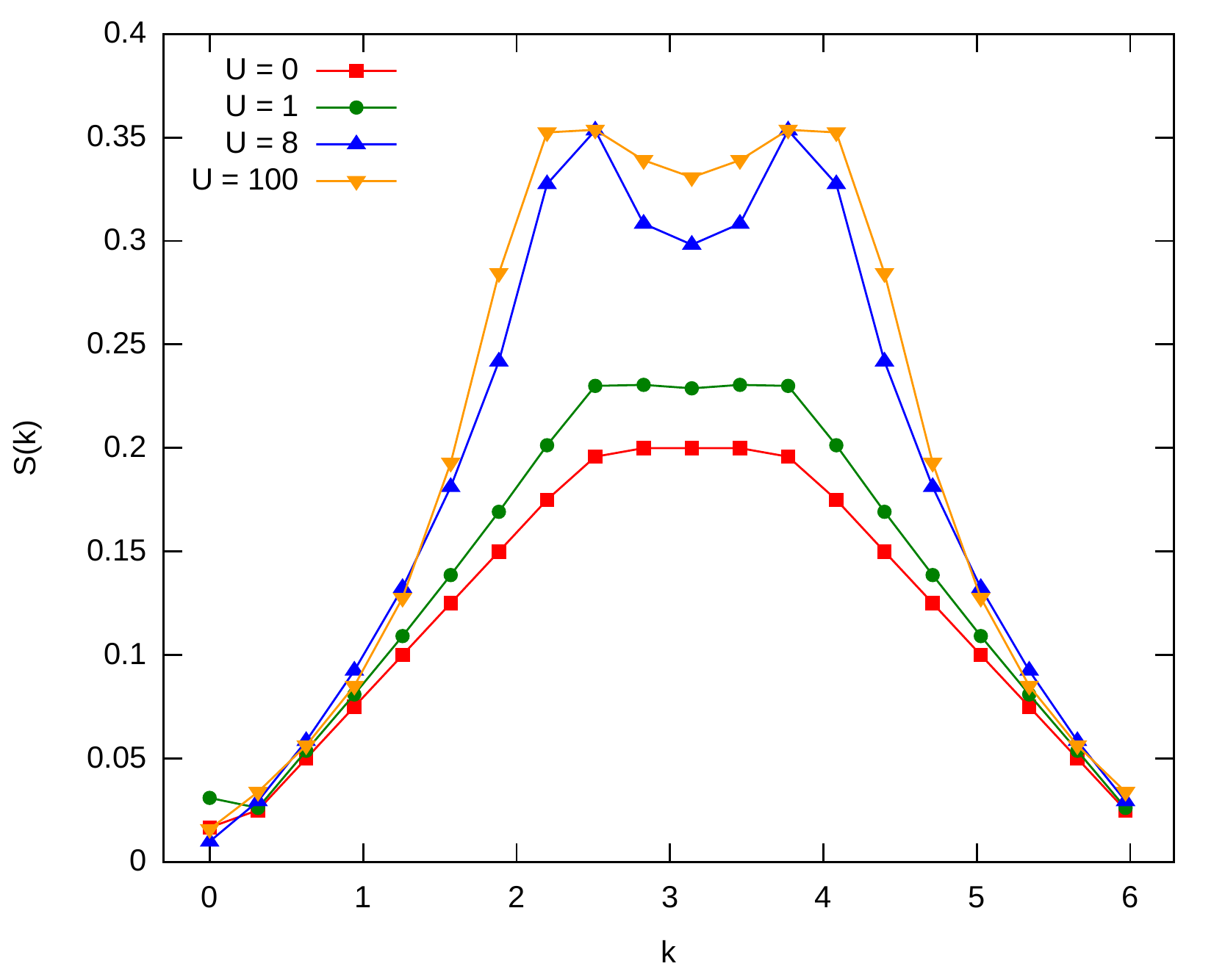}
\end{array}
$
\caption{\label{spincorr_0.4}Two-particle spin correlation function $S(k)$, as a function of momentum, for a $\frac{4}{10}$ filled lattice and various values of on-site repulsion $U$, using $\mathcal{IQG}$ (top) and $\mathcal{IQGT}$ (bottom) conditions.}
\end{figure}
\begin{figure}
\centering
$
\begin{array}{c}
\includegraphics[scale=0.7]{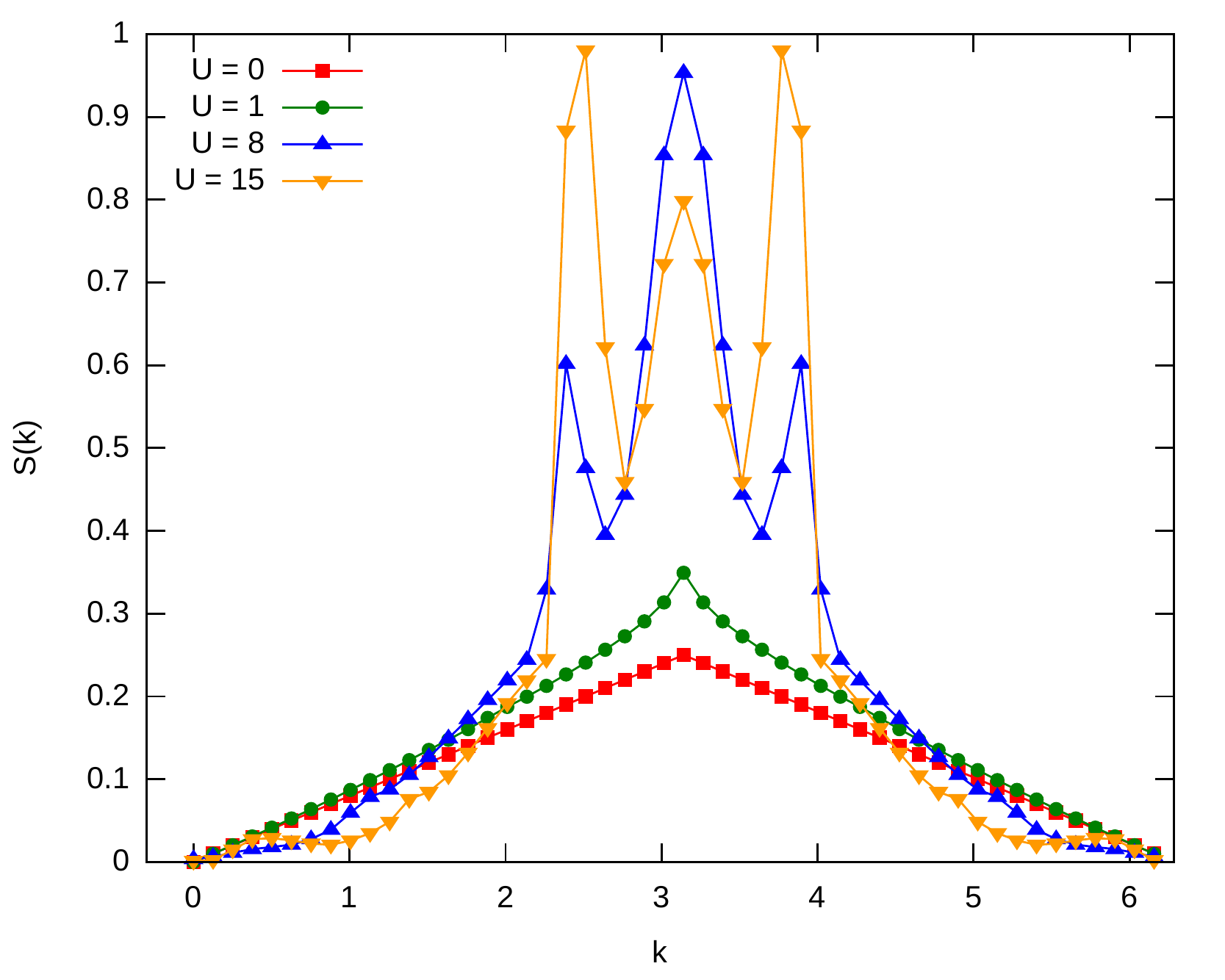}\\
\includegraphics[scale=0.7]{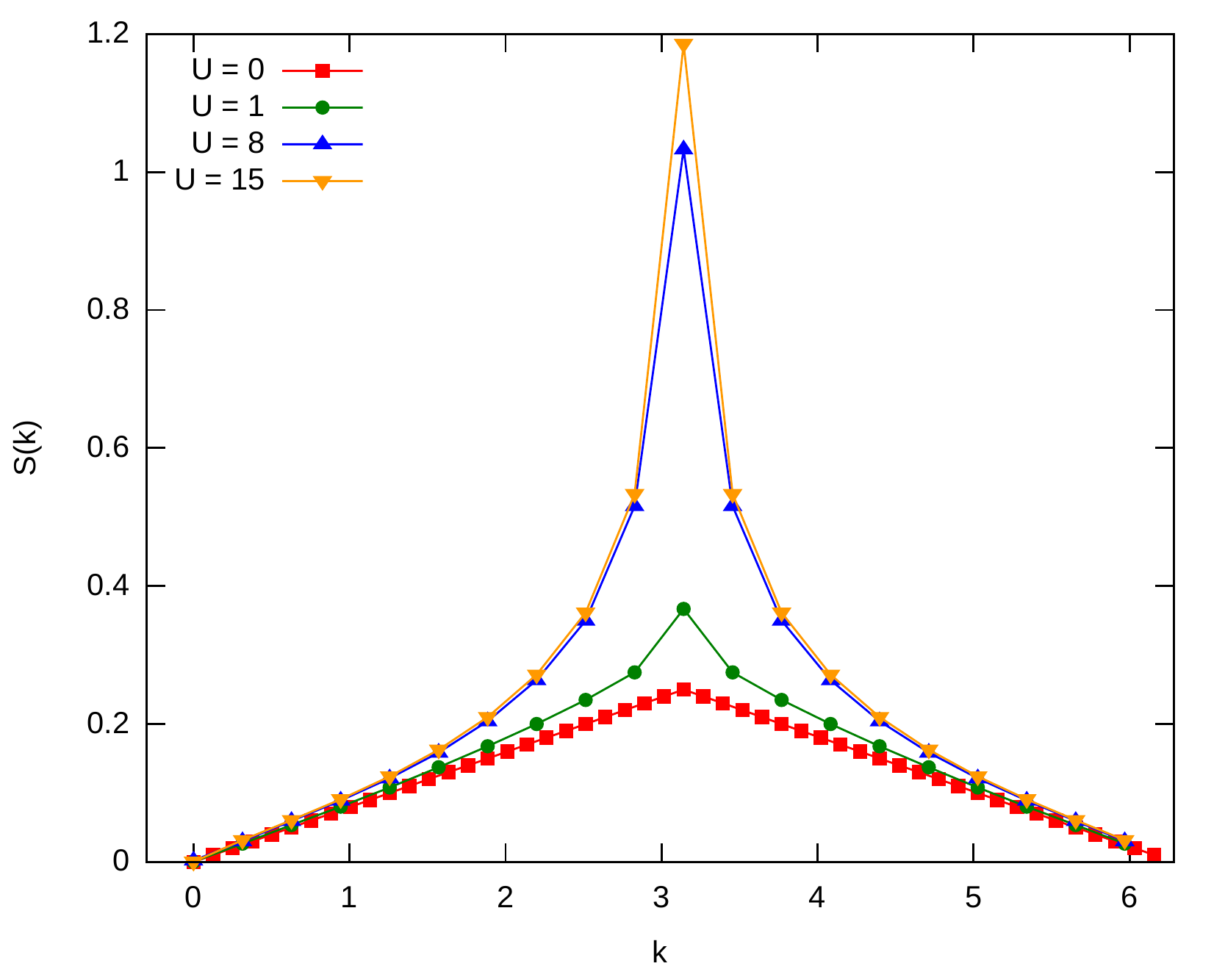}
\end{array}
$
\caption{\label{spincorr_0.5}Two-particle spin correlation function $S(k)$, as a function of momentum, for a half-filled lattice and various values of on-site repulsion $U$, using $\mathcal{IQG}$ (top) and $\mathcal{IQGT}$ (bottom) conditions.}
\end{figure}
The two-particle spin-correlation function defined as:
\begin{equation}
S(r) = \langle S_z^j S_z^{j+r}\rangle = \sum_{\sigma\sigma'} \sigma\sigma' \langle a^\dagger_{j\sigma}a_{j\sigma}a^\dagger_{j+r;\sigma'}a_{j+r;\sigma'} \rangle~,
\end{equation}
can be expressed as a function of the $\mathcal{G}$ matrix:
\begin{equation}
S(r) = \sum_{\sigma\sigma'}\sigma\sigma'\mathcal{G}(\Gamma)_{j\sigma j\sigma;(j+r)\sigma'(j+r)\sigma'}~.
\end{equation}
Written in terms of the spin-coupled $\mathcal{G}$ matrix, only the triplet $S=1$ part contributes:
\begin{equation}
S(r) = \frac{1}{2}\mathcal{G}(\Gamma)^1_{jj;(j+r)(j+r)}~.
\end{equation}
Fourier transforming $S(r)$ yields the momentum dependent spin-correlation function:
\begin{equation}
S(k) = \sum_r e^{ikr}S(r) = \frac{1}{2}\sum_{k_ak_b}\sum_{k_ck_d}\mathcal{G}^{1k}_{k_ak_b;k_ck_d}~.
\end{equation}
We have plotted this object in Figs.~\ref{spincorr_0.3}, \ref{spincorr_0.4} and \ref{spincorr_0.5} for $\frac{3}{10}$, $\frac{4}{10}$ and half filling respectively, using both $\mathcal{IQG}$ as $\mathcal{IQGT}$ conditions. Unsurprisingly, a good agreement is observed between the $\mathcal{IQG}$ and $\mathcal{IQGT}$ results for small values of $U$. At larger values of $U$, away from half filling, results are very poor with the $\mathcal{IQG}$ conditions, especially in the $\frac{4}{10}$-filled case, where the correlation function is wildly oscillating. More surprising, is that the spin-correlation function for the half-filled lattice in the large-$U$ limit is also incorrect in the $\mathcal{IQG}$ approximation. In the strong-correlation limit, for half-filling, the spin part of the Hubbard model is identical to the Heisenberg model \cite{ogata}, for which the spin-correlation function has a singularity at two times the Fermi momentum $2k_F=\pi$, which is exactly what we see in the $\mathcal{IQGT}$ results. Below half-filling the singularity in the large-$U$ limit splits and shifts to smaller values of $k$, as obserbed in the $\mathcal{IQGT}$ figures \ref{spincorr_0.3} and \ref{spincorr_0.4}. This is in agreement with the results in \cite{ogata} and \cite{sorella}. In conclusion we can say that the 2DM obtained with the $\mathcal{IQGT}$ conditions, correctly describes the physics that governs the spin-correlation function, whereas the $\mathcal{IQG}$ conditions do not. It is also important to note, that even though the $\mathcal{IQG}$ results for the energy are good for the half-filled lattice, the 2DM is flawed, because the spin-correlation function is not correctly described.
\section{Conclusion}
In this article we have studied the one-dimensional Hubbard model at various fillings using the v2DM method with both two- and three-index constraints. We have shown that it is possible to obtain a huge reduction in the computational cost of a basic matrix computation by exploiting all available symmetries, {\it i.e.} spin, translation invariance and space-inversion parity. Using this reduction it was possible to compare the computational scaling of different semidefinite programming algorithms with increasing lattice size. We found that, for this particular type of problem, the boundary point method outperforms interior point methods by several orders of magnitude. To gauge the quality of the variationally obtained 2DM we compared several ground-state properties to reference results. We found that, for half filling, the ground-state energy is well described by the two-index conditions. When moving away from half filling, however, we see that the three-index conditions are needed to obtain decent results. An explanation of why this happens has been the subject of a different article \cite{gutz_sdp}. The need for three-index constraints was even more obvious when we looked at the spin and charge correlation functions. It was also seen that, even though the energy was relatively correct for the half-filled lattice, the 2DM was flawed, because the spin correlation function was incorrect. This study shows that the exploitation of symmetry opens the possibility for a study of the two-dimensional Hubbard model for relevant lattice sizes. To obtain a decent accuracy, however, it will be necessary to include the three-index constraints, which is computationally hard. One way around this was set forward in \cite{gutz_sdp} with the use of lifting conditions \cite{mazziotti,hammond}.
\section{Acknowledgements}
We gratefully acknowledge financial support from FWO-Flanders and the research council of Ghent University. B.V., H.V.A., P.B., S.W. and D.V.N. are Members of the QCMM alliance Ghent-Brussels.
\bibliography{hubbard.bbl}
\end{document}